\documentclass[
    reprint,
    nofootinbib,
    amsmath,amssymb,
    aps,
    prd,
    floatfix,
]{revtex4-2}

\usepackage{tikz}
\usetikzlibrary{arrows.meta, shapes.geometric, positioning}
\usepackage{hhline}
\usepackage{graphicx}
\usepackage{dcolumn}
\usepackage{bm}
\usepackage[caption=false]{subfig}
\usepackage{makecell}
\usepackage{placeins}
\usepackage{varwidth}
\usepackage[breaklinks,colorlinks]{hyperref}
\hypersetup{%
,urlcolor=blue
,citecolor=blue
,linkcolor=blue
}

\newcommand{\half}{{\frac{1}{2}}}
\newcommand{\quart}{{\frac{1}{4}}}
\newcommand{\sgn}{\operatorname{sgn}}
\begin{document}

\preprint{APS/123-QED}

\title{Complete Classification of Domain Wall Solutions \\ in the $\mathbb{Z}_2$-symmetric 2HDM}

\author{Richard A. Battye}
    \email{richard.battye@manchester.ac.uk}
\author{Steven J. Cotterill}%
    \email{steven.cotterill@manchester.ac.uk}
\author{Adam K. Thomasson}%
    \email{adam.thomasson@manchester.ac.uk}

\affiliation{Department of Physics and Astronomy, The University of Manchester, Manchester, U.K.}

\date{\today}

\begin{abstract}
We present a complete classification of domain wall solutions in the two-Higgs Doublet Model (2HDM) with a global $\mathbb{Z}_2$ symmetry, categorised as superconducting, CP-violating, or neither, depending on the scalar particle masses and the ratio of the two Higgs doublets’ vacuum expectation values. We demonstrate that any domain wall solution can be reduced to depend on only six of the eight general field components, with further field reductions possible within different regions of the parameter space. Furthermore, we show that the superconducting solutions can be used to construct stable, current-carrying domain walls in two spatial dimensions. Similarly, the CP-violating solutions allow for two-dimensional configurations where CP symmetry is locally broken on the $\mathbb{Z}_2$-symmetric wall, which could provide an out-of-equilibrium environment for CP-violating processes to occur.
\end{abstract}

\maketitle


\section{Introduction}\label{sec:Intro}
Domain walls can form when a discrete symmetry is spontaneously broken at a phase transition in the Early Universe~\cite{Kibble:1976sj,Vilenkin278400}. These topological defects are generally viewed as problematic because, if absolutely stable, even a small population of domain walls would quickly dominate the Universe’s energy density unless the symmetry-breaking transition occurs at a sufficiently low energy scale~\cite{Zeldovich:1974uw}. However, if the discrete symmetry is only approximate – that is, if it is explicitly or softly broken – then any domain walls produced can decay on cosmologically acceptable timescales, rendering them compatible with observations~\cite{Larsson:1996sp}. Decaying domain walls can act as a potential source of gravitational waves (GWs), producing a stochastic background that may be observable with current or near-future experiments. As these metastable walls collapse under their own tension, they emit GWs with a characteristic spectral shape, typically peaking at frequencies determined by the wall energy scale and decay time \cite{BattyeGW, Hiramatsu_2010}.

Domain wall solutions in a simple $\mathbb{Z}_2$-symmetric model are well known~\cite{Vilenkin278400}. If that model is coupled to an additional field with an unbroken $U(1)$ symmetry, one finds superconducting wall solutions~\cite{Hodges:1988qg,Battye2008KV} which can form stable ring-like configurations in two dimensions known as “Kinky Vortons”~\cite{Battye2008KV}.

The two-Higgs Doublet Model (2HDM)~\cite{PhysRevD_8_1226} extends the Standard Model (SM) scalar sector by adding a second complex Higgs doublet (for a review see, for example, ref.~\cite{Branco_2012}). Such models are well-motivated and have been extensively studied – for example, they can introduce new sources of CP violation~\cite{Pilaftsis_1999, Keus_2016}, provide mechanisms for baryogenesis~\cite{Cohen_1993, Fromme2006, Grzadkowski_2009, Dorsch_2017} and even supply dark matter candidates~\cite{Grzadkowski_2009, Camargo_2019}. The 2HDM predicts the emergence of five physical scalar particles, $h$ and $H$ (CP-even neutral), $A$ (CP-odd neutral) and $H^\pm$ (charged). In the so-called \textit{alignment limit} (which we consider here), $h$ is identified with the observed SM Higgs boson. Phenomenological studies of the 2HDM are often performed in the $\mathbb{Z}_2$-symmetric variant as it can eliminate flavour changing neutral currents (FCNCs); this variant naturally leads to the formation of domain walls upon spontaneous symmetry breaking. Other discrete symmetries, CP1 and CP2, can also lead to the formation of domain walls~\cite{Battye2011VT,Chen2020, Eto20182HDMDW}. In all cases the discrete symmetries would need to be softly broken to avoid a domain wall over-closure problem~\cite{Battye2020CDW}.

Field-theoretic simulations of domain wall formation and evolution in the 2HDM have revealed some intriguing features. Notably, some domain wall solutions induce a non-zero photon mass at the centre of the wall~\cite{Battye2021SDW,Law_2022}. This local breaking of $U(1)_{\rm EM}$ has many potential implications~\cite{Battye_2021_photon,Sassi:2023cqp} that could be relevant if these effects last for a sufficient time - that is, if the soft-breaking scale is sufficiently low. More recent work~\cite{BATTYE2025139311} has shown that ring-like wall configurations – akin to the Kinky Vortons found in the simpler $\mathbb{Z}_2 \times U(1)$ model \cite{Battye2008KV, Battye2009FKV, Battye2009SKV} might also form in the 2HDM. These objects were not stable, but lasted much longer than would have been expected based on standard arguments of domain wall decay, with currents being observed on the walls. Kinky Vortons are $(2+1)$-dimensional analogues to Vortons~\cite{Witten:1984eb,DAVIS1989209,Lemperiere:2003yt,Battye:2008mm,Battye:2021sji,Battye:2021kbd}, where the superconducting string is replaced by a superconducting domain wall, allowing for studies of the stability within a lower-dimensional setting where greater dynamical range is available. The ultimate significance of such solutions for our (3+1)-dimensional universe, however, remains to be determined.

Inspired by the developments presented in refs.~\cite{Battye2011VT,Battye2021SDW,Law_2022,BATTYE2025139311,Sassi:2023cqp}, in this paper we fully categorise the parameter space of the $\mathbb{Z}_2$-symmetric 2HDM, identifying four possible subclasses of $\mathbb{Z}_2$ wall solutions. These have been categorised into: standard solutions, superconducting solutions\footnote{The $U(1)$ symmetry of electromagnetism is broken locally on the wall.}, CP-violating solutions\footnote{CP symmetry is locally broken on the wall, with the two vacua being CP preserving.} and superconducting \& CP-violating solutions\footnote{The solution is simultaneously Superconducting and CP-violating.}. We identify distinct regions of the parameter space where each subclass is the energy minimizing field configuration. We go on to demonstrate the existence of stable, current-carrying domain walls, formed from superconducting solutions, from which we suggest it might be possible to construct Kinky Vortons within this model. We also identify a novel two-dimensional solution in which a CP1 domain wall forms longitudinally along the core of a $\mathbb{Z}_2$ domain wall. This stable composite structure could provide a natural mechanism for out-of-equilibrium CP-violation. Our findings serve to successfully explain the observations of previously performed full dynamical simulations of this model.

\section{2HDM with $\mathbb{Z}_2$ Symmetry}\label{sec:theory}

\subsection{Formalism}
The Lagrangian density of the 2HDM (neglecting gauge fields and fermion couplings) can be written as
\begin{equation}
\mathcal{L} = (\partial^\mu \Phi_1)^\dagger (\partial_\mu \Phi_1) + (\partial^\mu \Phi_2)^\dagger (\partial_\mu \Phi_2) - V(\Phi_1, \Phi_2)\,,
\label{eq:2HDM_lag}
\end{equation}
where $\Phi_1$ and $\Phi_2$ are the two complex doublets. When the potential is restricted to be symmetric under the $\mathbb{Z}_2$ transformation $\Phi_1 \rightarrow \Phi_1,\, \Phi_2 \rightarrow - \Phi_2$ it takes the most general form of
\begin{align}
	V = & -\mu_{1}^2 (\Phi_1^\dag \Phi_1) - \mu_2^2 (\Phi_2^\dag \Phi_2) + \lambda_1 (\Phi_1^\dag \Phi_1)^2 + \lambda_2 (\Phi_2^\dag \Phi_2)^2 \cr 
    & + \lambda_3 	(\Phi_1^\dag \Phi_1)(\Phi_2^\dag \Phi_2) + (\lambda_4 - |\lambda_5|) \left [\mathrm{Re}(\Phi_1^\dag \Phi_2)  \right ]^2 \cr
    & + (\lambda_4 + |	\lambda_5|) \left [\mathrm{Im}(\Phi_1^\dag \Phi_2)  \right ]^2\,,
\label{eq:2HDM_Potential}
\end{align}
where the requirement of hermiticity demands that $\mu_1^2,\, \mu_2^2,\, \lambda_1,\, \lambda_2,\, \lambda_3,\, \lambda_4 \in \mathbb{R}$. In general $\lambda_5 \in \mathbb{C}$ but it can always be made real by a choice of basis \cite{Battye2011VT}, which we choose to represent in the potential using $|\lambda_5|$. It is often convenient to rewrite the potential in the so-called \textit{bi-linear field space} formalism \cite{Ivanov:2006yq, Ivanov_2008, Maniatis_2006},
\begin{equation}
    V = -\frac{1}{2}M_\mu R^\mu + \frac{1}{4}L_{\mu \nu} R^\mu R^\nu\,,
\end{equation}
where $M_\mu$ and $L_{\mu\nu}$ are constant coefficient matrices (their explicit forms can be found in, for example, ref.~\cite{Battye2011VT}) and $R^\mu$, which is invariant under a global electroweak (EW) transformation, is given by
\begin{equation}
R^\mu = \Phi^\dagger(\sigma^\mu \otimes \sigma^0)\Phi  = \begin{pmatrix} {\Phi_1^\dagger \Phi_1 + \Phi_2^\dagger \Phi_2} \\ {\Phi_1^\dagger \Phi_2 + \Phi_2^\dagger \Phi_1} \\ {-i[\Phi_1^\dagger \Phi_2 - \Phi_2^\dagger \Phi_1]} \\ {\Phi_1^\dagger \Phi_1 - \Phi_2^\dagger \Phi_2} \end{pmatrix}\,,
\label{eq:bi-linear_vec}
\end{equation}
where
$\Phi = \begin{pmatrix} \Phi_1 \\ \Phi_2\end{pmatrix}$. By introducing the $SU(2)_L$ invariant object, $\Phi_1^T i \sigma^2 \Phi_2$, $R^\mu$ can be promoted to a null 6-vector, $R^A$ for $A=0,..,5$, that incorporates charged degrees of freedom~\cite{Battye2011VT}. In particular, defining two additional components
\begin{eqnarray}
    R^4 &=& \Phi_1^T i \sigma^2 \Phi_2 - \Phi_2^\dagger i \sigma^2 \Phi_1^*\,, \cr
    R^5 &=& -i\left(\Phi_1^T i \sigma^2 \Phi_2 + \Phi_2^\dagger i \sigma^2 \Phi_1^*\right)\,,
\end{eqnarray}
one finds the useful identity $R^\mu R_\mu = R_4^2 + R_5^2$. The remaining degrees of freedom can be packaged into an $SU(2)_L$-vector $n^{a}=-\Phi^\dagger(\sigma^0\otimes\sigma^a)\Phi$ (with $a=1,2,3$), as discussed in ref.~\cite{Battye_2023}.

The two Higgs doublets of the model can be represented in various ways. In the \textit{linear representation}, we expand the fields in terms of eight real scalar components $\phi_i$,
\begin{equation}
    \Phi = \begin{pmatrix}\Phi_1 \\ \Phi_2 \end{pmatrix} = \begin{pmatrix}\phi_1 + i\phi_2 \\ \phi_3 + i\phi_4 \\ \phi_5 + i\phi_6 \\ \phi_7 + i\phi_8 \end{pmatrix}\,.
    \label{eq:lin_rep}
\end{equation}
Alternatively, one can describe the most general field configuration by applying an arbitrary EW rotation to a generic state containing only the degrees of freedom which affect the potential. In this \textit{general representation}, we write the field as an EW rotated vacuum state,
\begin{equation}
\Phi = \begin{pmatrix}\Phi_1 \\ \Phi_2 \end{pmatrix} = \frac{v_{\rm SM}}{\sqrt{2}}(\sigma^0 \otimes U) \begin{pmatrix}0 \\ f_1 \\ f_+ \\ f_2e^{i\xi} \end{pmatrix}\,,
\label{eq:gen_rep}
\end{equation}
where $v_{SM} = 246~\text{GeV}$ is the SM Higgs vacuum expectation value, and $U \in U(1)_Y \times SU(2)_L$ is a constant group element given, using the representation of ref.~\cite{Battye2021SDW}, by
\begin{equation}
U = e^{\half i\chi} \begin{pmatrix} \cos\half\gamma_1 e^{\half i\gamma_2} & \sin\half\gamma_1e^{\half i\gamma_3} \\ -\sin\half\gamma_1e^{-\half i\gamma_3} & \cos\half\gamma_1e^{-\half i\gamma_2}\end{pmatrix}\,.
\label{eq:EW_matrx}
\end{equation}
This generates an eight component field configuration, described by the \textit{vacuum manifold parameters} $f_1,\, f_+,\, f_2,\, \xi$, and the \textit{EW group parameters} $\chi,\, \gamma_1,\, \gamma_2,\, \gamma_3$. This configuration is \textit{charge breaking} in general, since a non-zero $f_+$ means the $U(1)$ symmetry of electromagnetism is broken. To enforce an electrically neutral vacuum, far from any defects, one can impose the condition $f_+ = 0$, which is equivalent to $R^\mu R_\mu = 0$ and is often referred to as the \textit{neutral vacuum condition}~\cite{Ivanov_2008}. As we and others have shown, this condition does not hold in the core of certain defect solutions, meaning those solutions carry a non-zero electromagnetic charge condensate.

There are useful relations connecting the bi-linear formalism to the representations above. For example, one can express the six-vector $R^A$ in terms of the general representation fields $(f_1,\, f_+,\, f_2,\, \xi,\, \chi)$ as
\begin{equation}
    R^A = \frac{v_{\rm SM}^2}{2}\begin{pmatrix}
        f_1^2 + f_+^2 + f_2^2 \\
        2f_1f_2\cos\xi \\
        2f_1f_2\sin\xi \\
        f_1^2 - f_+^2 - f_2^2 \\
        -2f_1f_+\cos\chi \\
        -2f_1f_+\sin\chi               
    \end{pmatrix}\,,
    \label{eq:RA_to_gen}
\end{equation}
and the EW group parameters can be related to the linear basis fields via
\begin{eqnarray}
    \gamma_1 &=& 2\arctan\left(\sqrt{\frac{\phi_1^2 + \phi_2^2}{\phi_3^2 + \phi_4^2}}\right)\,,\cr 
    \gamma_2 &=& \chi - 2\arctan\left(\frac{\phi_4}{\phi_3}\right)\,,\cr 
    \gamma_3 &=& 2\arctan\left(\frac{\phi_2}{\phi_1}\right) - \chi\,.
    \label{eq:gen_to_lin}
\end{eqnarray}

For our discussions in Sec.~\ref{sec:solution} it is useful to note a simplified form of the potential that emerges thanks to its $SU(2)_L \times U(1)_Y$ invariance,
\begin{align}
    V & = -\frac{\mu_1^2}{2}f_1^2-\frac{\mu_2^2}{2}(f_+^2 +f_2^2) + \frac{\lambda_1}{4}f_1^4 + \frac{\lambda_2}{4}(f_+^2 +f_2^2)^2
    \nonumber \\ & + \frac{\lambda_3}{4}f_1^2(f_+^2 +f_2^2) + \quart\left(\lambda_4 - |\lambda_5|\cos2\xi\right)f_1^2f_2^2\,,
    \label{eq:2HDM_Potential_2}
\end{align}
when we consider the general vacuum element of (\ref{eq:gen_rep}). We have set $v_{\rm SM} = 1$ in this expression for brevity, which we continue to do in all expressions related to the potential and energy densities from here on. We may consistently do this as it corresponds only to setting the energy and length scales, which we do in all numerical work presented: this is detailed in Appendix~\ref{sec:dim_rescaling}.

\subsection{Parameters}
The potential parameters of (\ref{eq:2HDM_Potential}) can be exchanged for a more physical set of quantities: the masses of the five scalar particles, $M_h,\, M_H,\, M_A$ and  $M_{H^\pm}$, the vacuum expectation value of the standard model, $v_{\rm SM}$, and the mixing angles between the CP-even and CP-odd fields of the model, $\alpha$ and $\beta$, with the ratio of the vacuum expectation values of the two field doublets given by $\tan\beta$, details of which can be found, for example, in ref.~\cite{Battye2021SDW}. To ensure that $h$ corresponds to the observed Higgs boson, we always work in the \textit{alignment limit} of $\cos(\alpha - \beta) = 1$ \cite{Battye2020CDW, Arbey_2018}. In this limit, the potential parameters can be written in simple closed forms in terms of the physical masses. For completeness, we list these relations here, 
\begin{align}
    \mu_1^2 & = \half M_h^2\,,  \quad \mu_2^2 = \half M_h^2\,,\cr
    \lambda_1 & = \frac{M_h^2 + M_H^2\tan^2\beta}{2v_{\rm{SM}}^2}\,, \quad \lambda_2 = \frac{M_h^2 + M_H^2\cot^2\beta}{2v_{\rm{SM}}^2}\,,\cr
    \lambda_3 & = \frac{(M^2_h - M^2_H) + 2M^2_{H^\pm}}{v_{\rm{SM}}^2}\,,\cr
    \lambda_4 & = \frac{M_A^2 - 2M^2_{H^\pm}}{v_{\rm{SM}}^2}\,, \quad |\lambda_5| = \frac{M_A^2}{v_{\rm{SM}}^2}.
    \label{eq:mass_mu_and_lambs}
\end{align}

Throughout this work, we refer to several specific parameter sets which have been chosen to illustrate different qualitative regimes of the model and to facilitate comparison with earlier work: they do not necessarily represent phenomenologically viable models. These parameter choices are summarized in Table~\ref{tab:Masses}. In each case we list the heavy scalar masses $M_H,\, M_A,\, M_{H^\pm}$ and $\tan\beta$ (with alignment limit assumed). The values of $M_h = 125\,{\rm GeV}$ and $v_{SM} = 246\,{\rm GeV}$ are fixed by experiment \cite{Workman:2022ynf}, however as detailed in Appendix~\ref{sec:dim_rescaling} we set the energy and length scales in our numerical work by rescaling the parameters such that these are both set to unity. For most investigations in this work, for example the parameter scans of Sec.~\ref{sec:solution}, we maintain that $\tan\beta = 0.85$ unless otherwise stated, but we show that the conclusions drawn are not limited only to this case.

\begin{table}
    \centering
    \small
    \setlength{\tabcolsep}{5pt}  
    \renewcommand{\arraystretch}{1.2}  
    \begin{tabular}{| c | c | c | c | c |}
        \hline
        Parameter Set & $M_H$ & $M_A$ & $M_{H^\pm}$ & $\tan\beta$ \\
        \hline
        \multicolumn{5}{|c|}{} \\[-3ex]  
        \hline
		      A & 200 & 200 & 200 & 0.85 \\
            B & 600 & 200 & 300 & 0.85 \\
            C & 600 & 300 & 300 & 0.85 \\
            D & 600 & 300 & 200 & 0.85 \\
            E & 600 & 400 & 300 & 0.85 \\
            F & 600 & 300 & 400 & 0.85 \\
            G & 750 & 900 & 125 & 0.35 \\
            H & 500 & 125 & 250 & 0.85 \\
		\hline
	\end{tabular}
    \caption{Table of mass parametrisations used in this work, with masses given in $\rm GeV$. All parametrisations used assume $M_h = 125\,{\rm GeV},\, v_{\rm SM}= 246\,{\rm GeV}$ and $\cos(\alpha - \beta) = 1$. Parameter set A was that used in ref.~\cite{Battye2021SDW}.}
    \label{tab:Masses}
\end{table}

\subsection{Field Configuration Reduction}\label{sec:field_reduction}
While the general field ansatz of (\ref{eq:gen_rep}) involves eight continuous field components, finite energy arguments can reduce the number of independent degrees of freedom needed to describe a domain wall. We introduce a convenient representation of the EW rotation matrix of (\ref{eq:EW_matrx}) that leads to a symmetric form for this simplification. 

However, the parameter reduction we discuss is not unique to a particular representation and can be understood by looking at the internal structure of $SU(2)$. The group admits the decomposition $U = U_1 U_2 = e^{i |u| \hat u^a \sigma^a} e^{i v^b \sigma^b}$ with $\hat u^a$ a chosen constant direction and $v^b$ restricted to lie in the plane spanned by $\hat{u}^a$ and one of its two perpendicular directions, where $U_1 \in U(1)_{\hat u^a} \subset SU(2)$ and $U_2 \in SU(2)/U(1)_{\hat u^a}$. The group parameters only enter into the energy density through the gradients, therefore we need only consider $U^\dagger \partial_x U = U_2^\dagger U_1^\dagger (\partial_x U_1) U_2 + U_2^\dagger \partial_x U_2$, where $U_1^\dagger \partial_x U_1 = i e^{-i |u| \hat u^a \sigma^a} (\partial_x |u|) \hat u^b \sigma^b e^{i |u| \hat u^c \sigma^c} = i (\partial_x |u|) \hat u^a \sigma^a$, thanks to the fixed $\hat u^a$, such that only the derivatives of $|u|$ enter the energy density. Similarly, the $U(1)_Y$ gauge parameter enters only through its derivatives, in complete analogy with the internal $U(1)$ subgroup of $SU(2)$, whereas there is an explicit dependence on the two components of $v^b$ as well as on their derivatives. As such, only two of the group parameters appear explicitly in the energy density.

Our parametrisation introduces three new angles $\eta_1,\,\eta_2,\,\eta_3$ defined by
\begin{equation}
    \eta_1 = \gamma_2 - \gamma_3\,,\quad \eta_2 = \chi+\gamma_3\,, \quad \eta_3 = \chi-\gamma_3\,,
\end{equation}
in terms of which we can rewrite the group element as a product of two matrices,
\begin{equation}
    U = \begin{pmatrix} e^{\half i \eta_2} & 0 \\ 0 & e^{\half i \eta_3} \end{pmatrix}\begin{pmatrix}
        \cos\half\gamma_1 e^{\half i \eta_1} & \sin\half\gamma_1 \\ -\sin\half\gamma_1 & \cos\half\gamma_1 e^{-\half i \eta_1}\end{pmatrix}\,.
\label{eq:gen_sol_matrix}
\end{equation}
This representation is advantageous when considering static, finite-energy domain wall solutions. In particular, by demanding that the energy density of the wall remains localized, one finds that $\eta_2$ and $\eta_3$ can be expressed in terms of the remaining six components in the wall configuration and as such explicitly eliminated; we will show later that they in fact converge to zero in all numerical solutions we have found. As a result, a general domain wall solution can be described using only six independent functions of $x$: namely $f_1$, $f_+$, $f_2$, $\xi$, $\gamma_1$, and $\eta_1$. In specific regions of parameter space, further simplifications occur – for example, some of these functions vanish, yielding special subclasses of solution, as we will explore in Sec.~\ref{sec:solution}.

To illustrate our proposed field reduction, we can write the one-dimensional energy density for the general field configuration and examine its structure. The one dimensional energy density for this configuration is given by
\begin{widetext}
\begin{eqnarray}
    \mathcal{E} & = & \half(\partial_x f_1)^2 + \half(\partial_x f_+)^2 + \half(\partial_x f_2)^2 + \half f_2^2(\partial_x \xi)^2 + \frac{1}{8}(f_1^2+f_+^2+f_2^2) \Big[(\partial_x\gamma_1)^2 + c_1^2 (\partial_x\eta_1)^2 \Big] \cr
    & + & \half c_\xi\Big[(f_2\partial_x f_+ - f_+\partial_x f_2)\partial_x\gamma_1 + f_+f_2s_1c_1\partial_x\xi\partial_x\eta_1\Big] + \half s_\xi\Big[(f_+\partial_x f_2 - f_2\partial_x f_+)s_1c_1\partial_x\eta_1 + f_+f_2\partial_x\xi\partial_x\gamma_1\Big] \cr
    & - & \half f_2^2c_1^2\partial_x\xi\partial_x\eta_1 + \mathcal{E}_{\eta_2} + \mathcal{E}_{\eta_3} + V\,,\cr
\text{where}\cr
    \mathcal{E}_{\eta_2} & = &\frac{\partial_x\eta_2}{2}\Bigg[f_2^2s_1^2 \partial_x\xi + \half f_+^2c_1^2 \partial_x\eta_1  + \half f_+f_2s_\xi\partial_x\gamma_1  + s_1c_1 \bigg\{f_+f_2c_\xi\bigg(\partial_x\xi +\frac{\partial_x\eta_1}{2}\bigg) +  (f_+\partial_x f_2 - f_2\partial_x f_+)s_\xi\bigg\}\Bigg] \cr
    & + & \frac{(\partial_x \eta_2)^2}{8}\Bigg[(f_1^2+f_2^2)s_1^2 + f_+^2c_1^2  + 2f_+f_2s_1c_1c_\xi\Bigg]\,, \cr
    \mathcal{E}_{\eta_3} & = & \frac{\partial_x\eta_3}{2} \Bigg[f_2^2c_1^2 \partial_x\xi-\half (f_1^2 + f_2^2)c_1^2 \partial_x\eta_1 + \half f_+f_2s_\xi\partial_x\gamma_1 - s_1c_1 \bigg\{f_+f_2c_\xi\bigg(\partial_x\xi -\frac{\partial_x\eta_1}{2}\bigg) + (f_+\partial_x f_2 - f_2\partial_x f_+)s_\xi\bigg\}\Bigg] \cr
    & + & \frac{(\partial_x \eta_3)^2}{8}\Bigg[(f_1^2+f_2^2)c_1^2 + s_1^2 f_+^2 - 2f_+f_2s_1c_1c_\xi\Bigg]\,,
    \label{eq:gen_energy}
\end{eqnarray}
\end{widetext}
where we have defined $s_1 = \sin\half\gamma_1$, $c_1 = \cos\half\gamma_1$, $s_\xi = \sin\left(\xi - \half\eta_1\right)$, $c_\xi = \cos\left(\xi - \half\eta_1\right)$ for brevity. The two EW group parameters $\eta_2$ and $\eta_3$ do not appear explicitly, but only implicitly through their first derivatives. Taking the resulting equation of motion for $\eta_2$,
\begin{widetext}
\begin{eqnarray}
    &&\partial_x \Bigg\{\half\partial_x \eta_2\bigg[ s_1^2(f_1^2+f_2^2) +  c_1^2 f_+^2 + 2s_1c_1f_+f_2c_\xi\bigg] + \bigg[s_1^2 f_2^2\partial_x\xi + \half c_1^2 f_+^2\partial_x\eta_1 + \half f_+f_2s_\xi\partial_x\gamma_1 \cr 
    && + s_1c_1 \bigg\{f_+f_2c_\xi\left(\partial_x\xi +\half \partial_x\eta_1\right) + \left(f_+\partial_xf_2 - f_2\partial_xf_+\right)s_\xi\bigg\}\bigg] \Bigg\} = 0\,,
\end{eqnarray}
\end{widetext}
we see that the contents of the outermost brackets must therefore equal a constant. For finite energy solutions the derivatives of all functions must go to zero far from the wall, and as such the constant must equal zero. We may also use the same argument for $\eta_3$, such that we obtain the two expressions
\begin{widetext}
\begin{equation}
\partial_x\eta_2 = - \frac{s_1^2 f_2^2\partial_x\xi + \half c_1^2 f_+^2\partial_x\eta_1
         + \half f_+f_2 s_\xi\partial_x\gamma_1
         + s_1c_1 \bigg\{f_+f_2c_\xi\left(\partial_x\xi +\half \partial_x\eta_1\right)
         + \left(f_+\partial_xf_2 - f_2\partial_xf_+\right)s_\xi\bigg\}}
         {\half s_1^2(f_1^2+f_2^2) + \half c_1^2 f_+^2 + s_1c_1f_+f_2c_\xi}\,,
    \label{eq:deta2}
\end{equation}
\begin{equation}
\partial_x\eta_3 = -\frac{c_1^2 f_2^2\partial_x\xi-\half c_1^2 (f_1^2 + f_2^2)\partial_x\eta_1 + \half f_+f_2s_\xi\partial_x\gamma_1 - s_1c_1 \bigg\{f_+f_2c_\xi\left(\partial_x\xi -\half \partial_x\eta_1\right) + \left(f_+\partial_xf_2 - f_2\partial_xf_+\right)s_\xi\bigg\}}{\half c_1^2(f_1^2+f_2^2) + \half s_1^2 f_+^2 - s_1c_1f_+f_2c_\xi}\,,       
\end{equation}
\end{widetext}
which may be substituted into the remaining six equations of motion, or back into the energy density to eliminate the dependence on $\eta_2$ and $\eta_3$ explicitly. This results in any domain wall solution within the model depending on only six field components, $f_1,\, f_+,\, f_2,\, \xi,\, \gamma_1 \text{ and } \eta_1$, from which six consistent equations of motion may be derived.\footnote{In the language of classical mechanics, $\eta_2$ and $\eta_3$ are \textit{cyclic} (or \textit{ignorable}) generalized coordinates: the one–dimensional Lagrangian density (or static energy density) depends on them only through their first derivatives, so their conjugate momenta are conserved. The finite energy conditions described above require these conserved momenta to vanish identically. In this special case, the cyclic coordinates could also be eliminated directly by substitution into the Lagrangian also, without altering the equations of motion for the remaining fields. However, in the generic case, where the associated conjugate momenta are non-zero, direct substitution into the Lagrangian changes the variational problem and yields incorrect reduced equations of motion; the consistent procedure is to perform a partial Legendre transform in the cyclic coordinates (Routhian reduction) or to work in the Hamiltonian formalism \cite{Goldstein2002,Landau1976,Arnold1989}, as we have done here by using the one-dimensional energy density.}

In Sec.~\ref{sec:solution} we use this reduced field description to classify all possible domain wall solutions and identify the parameter conditions under which each occurs.

\section{Motivation}\label{sec:motivation}
\subsection{Full Dynamical Simulations}\label{sec:RIC_overview}
Numerical simulations of the 2HDM from random initial conditions (RIC) indicate~\cite{Battye2021SDW} that the domain wall solution of ref.~\cite{Battye2011VT} does not naturally form in simulations of the same parameters (parameter set A). Instead, the walls in the simulation develop a non-zero $R^\mu R_\mu$ (i.e. the walls become superconducting). Moreover, although the bi-linear components $R^0$, $R^1$ and $R^3$ behaved as in the known solution ($R^1$ changes sign across the wall and $R^0, R^3$ have the expected extrema), the simulations showed a notable discrepancy: $R^2$ is non-zero on the domain walls, whereas in the ref.~\cite{Battye2011VT} solution $R^2$ vanishes.\footnote{Ref.~\cite{Battye2021SDW} did not comment on the non-zero $R^2$ observed on the walls, it was first drawn to attention in our recent work of ref.~\cite{BATTYE2025139311}; the present work is the first to analyse this feature in the 2HDM} In other words, the domain walls that form dynamically tend to excite fields that are found to be zero in the solution of ref.~\cite{Battye2011VT}.

\begin{figure*}
    \centering
    \subfloat[A]{
        \begin{minipage}[t]{0.22\textwidth}
            \centering
            \includegraphics[width=\columnwidth]{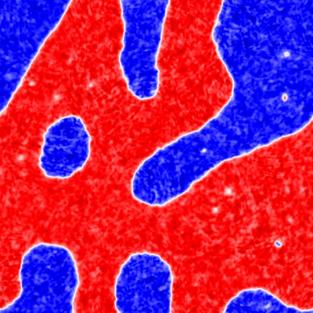}\vspace{4pt} \\
            \includegraphics[width=\columnwidth]{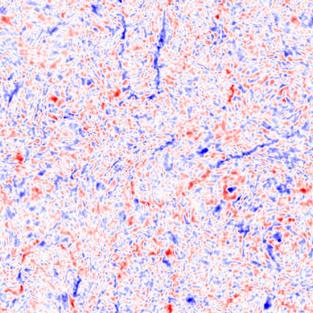}\vspace{4pt} \\
            \includegraphics[width=\columnwidth]{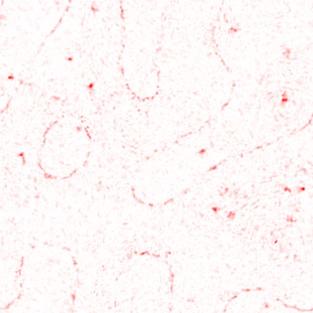}
            \label{fig:RIC_222}
        \end{minipage}
    }
    \subfloat[B]{
        \begin{minipage}[t]{0.22\textwidth}
            \centering
            \includegraphics[width=\columnwidth]{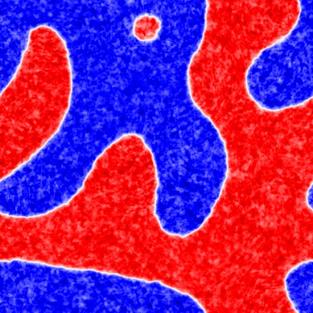}\vspace{4pt} \\
            \includegraphics[width=\columnwidth]{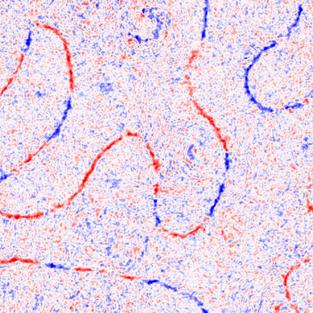}\vspace{4pt} \\
            \includegraphics[width=\columnwidth]{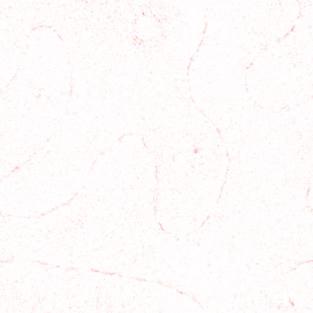}
            \label{fig:RIC_623}
        \end{minipage}
    }
    \subfloat[C]{
        \begin{minipage}[t]{0.22\textwidth}
            \centering
            \includegraphics[width=\columnwidth]{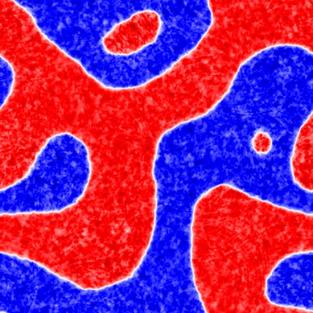}\vspace{4pt} \\
            \includegraphics[width=\columnwidth]{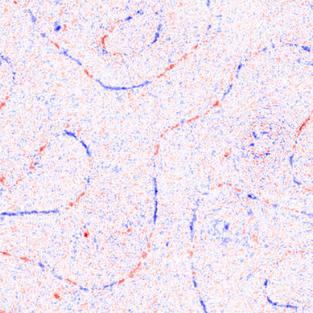}\vspace{4pt} \\
            \includegraphics[width=\columnwidth]{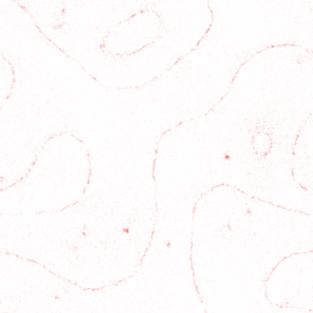}
            \label{fig:RIC_633}
        \end{minipage}
    }
    \subfloat[D]{
        \begin{minipage}[t]{0.22\textwidth}
            \centering
            \includegraphics[width=\columnwidth]{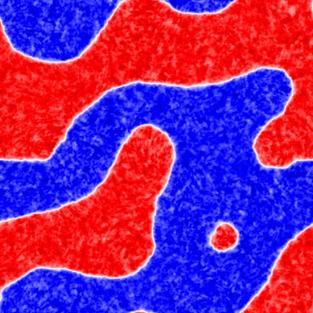}\vspace{4pt} \\
            \includegraphics[width=\columnwidth]{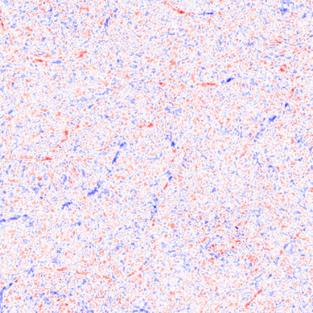}\vspace{4pt} \\
            \includegraphics[width=\columnwidth]{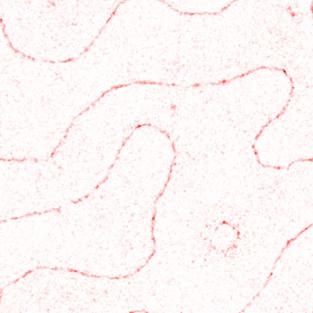}
            \label{fig:RIC_632}
        \end{minipage}
    }
    \subfloat{
        \begin{minipage}[t]{0.0322\textwidth}
            \centering
            \includegraphics[width=\columnwidth]{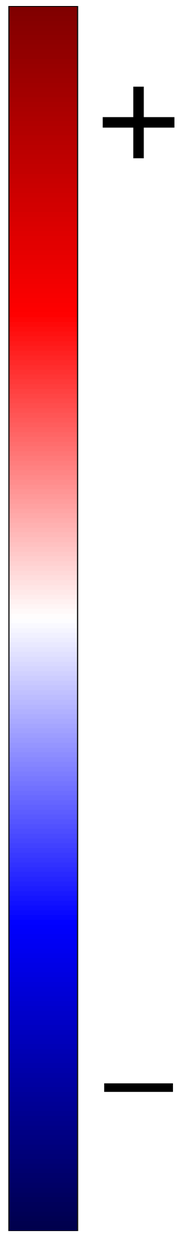}\vspace{4pt} \\
            \includegraphics[width=\columnwidth]{RICcb.png}\vspace{4pt} \\
            \includegraphics[width=\columnwidth]{RICcb.png}
        \end{minipage}
    }
    \caption{$(2+1)$-dimensional simulations of the 2HDM with a $\mathbb{Z}_2$-symmetric potential for different mass parametrisations (A, B, C and D). Shown are the bi-linear components $R^1$ (top), $R^2$ (middle), and $R^\mu R_\mu$ (bottom). Colour mappings of each individual component are normalized across all parameter sets, with each component type having a separate scale.}
    \label{fig:RIC_overview}
\end{figure*}

Fig.~\ref{fig:RIC_overview} illustrates these findings by showing the spatial profiles of $R^1$, $R^2$ and $R^\mu R_\mu$ in a set of $(2+1)$-dimensional field simulations from RIC, for four different parameter sets. Fig.~\ref{fig:RIC_222} reproduces the result for parameter set A (used in ref.~\cite{Battye2021SDW}) showing that both $R^2$ and $R^\mu R_\mu$ are mildly localized on the domain walls. We also show corresponding simulations for three alternative parameter sets (B, C and D) to highlight how the wall structure varies with the mass hierarchy; these specific choices will be analysed in detail in Sec.~\ref{sec:solution}, but here we note the key differences. In all three additional parameter sets either $R^2$ and/or $R^\mu R_\mu$ become large on the walls. When $M_A = M_{H^\pm}$ (C), both quantities are moderately elevated on the walls. However, when $M_{H^\pm} < M_A$ (D), the simulations show a significantly larger $R^\mu R_\mu$ condensate and a much smaller $R_2$ on the walls. This trend reverses for $M_{H^\pm} > M_A$ (B), where a prominent $R^2$ accompanies a suppressed $R^\mu R_\mu$.
Note that where $R^\mu R_\mu$ is found to be non-zero on domain walls a ``winding" is observed in both the extended components of $R^4$ and $R^5$, although is not depicted here. These enhanced features correlate directly with the expected minimum-energy solutions for each parameter set, as we will demonstrate in Sec.~\ref{sec:solution}. In contrast, the original, parameter set A, solution from ref.~\cite{Battye2011VT} (which we confirm to be of minimum-energy) exhibits neither feature found to occur in dynamical simulations.

Some definitions used throughout this work which are important to highlight are:
\begin{itemize}
    \item \textbf{Minimum-Energy Solution}: The lowest possible energy solution for a given set of parameters $(M_H,\, M_A,\, M_{H^\pm},\, \tan\beta)$, where all general field components are allowed to vary and homogeneous Neumann boundary conditions are imposed.
    \item \textbf{Naturally Bounded Solution}: Solutions where homogenous Neumann boundary conditions are imposed on all field variables. These boundary conditions are \textit{natural} for domain wall solutions such that there is no variation of the energy of the fields in the vacuum. These solutions are not necessarily of minimum-energy, for example when a reduced field ansatz is used.
    \item \textbf{Standard Solution}: Domain wall solution found in ref.~\cite{Battye2011VT}, where $f_+ = \xi = \gamma_1 = \eta_1 = \eta_2 =\eta_3 \equiv 0$, described only by the general field variables $f_1$ and $f_2$.
\end{itemize}

We will show in Sec.~\ref{sec:solution} that the parameter space of the model divides into distinct regions, characterised by whether the minimum-energy wall solution supports non-zero $R^\mu R_\mu$, non-zero $R^2$, both, or neither. These solutions can be described by reduced ans\"atze, obtained by restricting the general field configuration of Sec.~\ref{sec:field_reduction}, thereby greatly simplifying the analysis of each region. However, as we have detailed above with Fig.~\ref{fig:RIC_overview}, simulations reveal that the fields do not fully relax to these minimum-energy configurations: the walls exhibit small but non-zero values of $R^\mu R_\mu$ and/or $R^2$ even when the corresponding minimum-energy state does not. These apparent ``excitations'' of the walls can be traced back to the fact that the fields emerge from random initial conditions, as expected in a phase transition, leading to relative EW transformations between neighbouring vacua. These field configurations have been previously discussed in refs.~\cite{Battye2021SDW, Law_2022, Sassi:2023cqp}, but their role has not yet been systematically characterised. We first aim to address these relative EW transformations to explain the departures from the minimum-energy state observed in full dynamical simulations.

It is important to note here that a non-zero value of $R^2$ on a domain wall requires an interpolating profile of the phase $\xi$ between $0$ and $\pi$ across the wall, and thus a non-zero value of $f_2$ at its centre. This follows directly from the structure of $R^2$ in (\ref{eq:RA_to_gen}). The field component $f_2e^{i\xi}$ can be written in two equivalent ways depending on the choice of domain, $f_2 \in\mathbb{R}_{\geq 0},\, \xi \in(-\pi, \pi]$ or $f_2 \in\mathbb{R},\, \xi \in(-\pi/2, \pi/2]$. For solutions with $f_2(0) = 0$, the wall may be described either as an interpolation from negative to positive values of $f_2$ with $\xi\equiv0$ or, equivalently, as a discontinuity in $\xi$ between $0$ and $\pi$ and $f_2\geq0$ -- both descriptions coincide smoothly in the limit $f_2 \to 0$. For solutions with $f_2(0) \neq 0$, the wall can be considered to interpolate smoothly in $\xi$ with $f_2>0$, producing a non-zero $R^2$, or as an equivalent description that involves discontinuities in both $\xi$ and $f_2$ at the centre. In both cases, we choose the description without discontinuities as this is both simpler and more convenient for numerical work, although this does mean that we have chosen different domains for the two scenarios. We will often refer to the former type of solution by stating that ``the wall is in $f_2$" and for the latter ``the wall is in $\xi$". Previous studies have only considered walls in $f_2$.

\subsection{Relative Electroweak Transformations of the Vacua}\label{sec:EW_Rel_Trans}
As mentioned above, it has been explained \cite{Battye2021SDW} that one can generally perform a relative EW transformation of the vacuum upon either side of the domain wall. In practice, this means that the vacuum expectation values on either side of the wall need not be in the same electroweak gauge - they can differ by a constant $U(1)_Y \times SU(2)_L$ transformation. This situation was first explored in ref.~\cite{Battye2021SDW} as it would be a natural occurrence in full dynamical simulations from RIC.

Such a field configuration takes the form
\begin{eqnarray}
\Phi(-\infty) &=&\frac{v_{\rm SM}}{\sqrt{2}}\begin{pmatrix} 0 \\ {\bar f}_1 \\ 0 \\ - {\bar f}_2 \end{pmatrix}\,,\cr
\Phi(+\infty) &=& \frac{v_{\rm SM}}{\sqrt{2}}\big(\sigma^0 \otimes U\big)\begin{pmatrix} 0 \\ {\bar f}_1 \\ 0 \\ {\bar f}_2\end{pmatrix}\,,
\label{eq:EW_trans_boundaries}
\end{eqnarray}
where $U\equiv U(\chi,\gamma_1,\gamma_2,\gamma_3)$ is a constant $U(1)_Y \times SU(2)_L$ rotation matrix as given in (\ref{eq:EW_matrx}) and $({\bar f}_1, {\bar f}_2)=(\cos\beta,\sin\beta)$ are the vacuum expectation values for the two fields, with the sign of the lower component changing as in a $\mathbb{Z}_2$ domain wall. Ref.~\cite{Battye2021SDW} only studied a small number of different relative rotations, whereas ref.~\cite{Law_2022} considered specific lines within the parameter space using a different but equivalent EW representation. Here we report a systematic study of the solutions as functions of $\chi,\,\gamma_1,\,\gamma_2$ and $\gamma_3$, completing the delineation of this issue.

Firstly, we note that we can express
\begin{equation}
 \big(\sigma^0 \otimes U\big)\begin{pmatrix} 0 \\ {\bar f}_1 \\ 0 \\ {\bar f}_2\end{pmatrix} =  \begin{pmatrix} {\bar f}_1\sin\half\gamma_1 e^{\half i\theta_2} \\ {\bar f}_1\cos\half\gamma_1 e^{\half i\theta_1} \\ {\bar f}_2\sin\half\gamma_1 e^{\half i\theta_2} \\ {\bar f}_2\cos\half\gamma_1 e^{\half\theta_1}\end{pmatrix}\,,
\label{eq:EW_trans_elim}
\end{equation}
where $\theta_1 = \chi - \gamma_2= \eta_3 - \eta_1$ and $\theta_2 = \chi + \gamma_3 =\eta_2$. This has one less EW degree of freedom, than the general element (\ref{eq:EW_matrx}), due to the demand of the unbroken $U(1)_{\rm EM}$ symmetry in the vacuum. Importantly, some of the remaining continuous parameters are physically redundant, as an additional global EW rotation can be applied that does not alter $\Phi(-\infty)$ but does alter $\Phi(+\infty)$. This transformation is the unbroken $U(1)_{\rm EM}$ symmetry at negative infinity, but not at positive infinity, and takes the form $\sigma^0\otimes{\bar U}$ where
\begin{equation}
{\bar U} = \begin{pmatrix}
	e^{-\half i\theta_2} & 0 \\
	0 & 1
	\end{pmatrix}\,,
\label{eq:theta2_elim_trans}
\end{equation}
which eliminates the $\theta_2$ dependence of $\Phi(+\infty)$,
\begin{equation}
  \Phi(+\infty) = \frac{v_{\rm SM}}{\sqrt{2}}\begin{pmatrix} {\bar f}_1\sin\half\gamma_1 \\ {\bar f}_1\cos\half\gamma_1 e^{\half i\theta_1} \\  {\bar f}_2\sin\half\gamma_1 \\ {\bar f}_2\cos\half\gamma_1e^{\half i\theta_1}\end{pmatrix}\,.
\label{eq:EW_trans_red_boundaries}
\end{equation}

\begin{figure}
\centering
\includegraphics[width=\columnwidth]{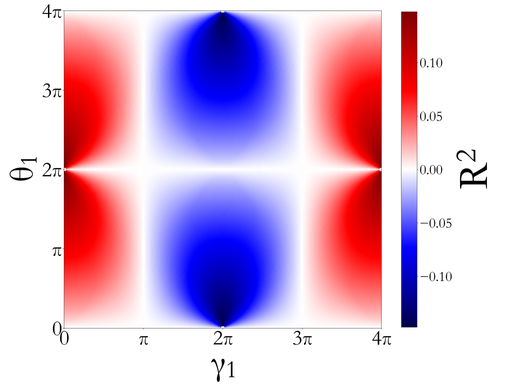}
\vspace{5pt}
\label{fig:EW_param_space_R2}
\includegraphics[width=\columnwidth]{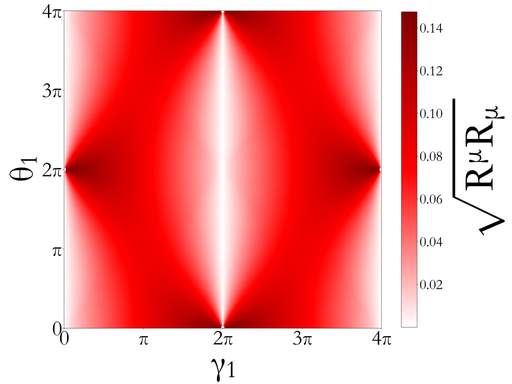}
\vspace{5pt}
\label{fig:EW_param_space_RmuRmu}
\includegraphics[width=\columnwidth]{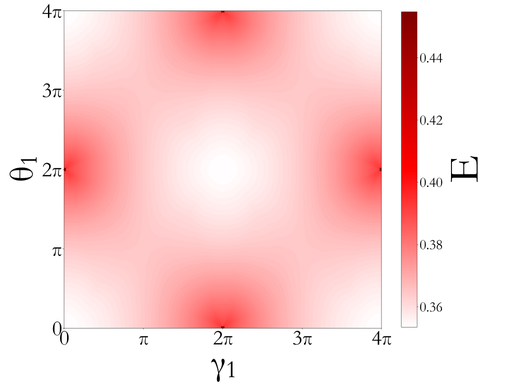}
\label{fig:EW_param_space_energy}
\caption{Values of $R^2$ and $\sqrt{R^\mu R_\mu}$ at the centre of fixed boundary domain wall solutions, and the total energy, E, as functions of $\gamma_1$ and $\theta_1$ in parameter set A, for relatively EW rotated vacua. The rotations were varied in increments of $\pi/128$ in the range $[0, 4\pi]$, such that $512^2$ different solutions were computed.}
\label{fig:EW_param_space}
\end{figure}

Choosing parameter set A for this study, we solved for kink solutions by identifying the far vacuum values given by (\ref{eq:EW_trans_red_boundaries}) with those in the linear representation and evolving the equations of motion within the latter. We used the \textit{Approximate Treatment} detailed in Appendix~\ref{sec:numericals_kinks} for this investigation, with numerical settings of $n_x=300,\, \Delta x=0.1,\, \delta=10^{-5}$ for computational ease, as such a treatment and resolution is acceptable when working in the linear representation: the field equations are only softly coupled. As is the nature of this type of solution, we impose fixed boundary conditions on the fields, varying the two angles $\gamma_1$ and $\theta_1$ between $0 \text{ and } 4\pi$ for full coverage of the field configuration. The values of $R^2$ and $R^\mu R_\mu$ at the centre of the kink solutions are shown in Fig.~\ref{fig:EW_param_space} across this relative transformation space, with the energy of the solutions also included. This analysis shows that only under zero or an exact $\gamma_1 = \theta_1 = 2\pi$ rotation does the field configuration correspond to that of the minimum-energy solution for this parameter set, and as such any small relative rotation between the two vacua in simulations will result in altered wall properties, ie. $R^2 \neq 0$ and/or $R^\mu R_\mu \neq 0$. This fully explains the features of the simulation depicted in Fig.~\ref{fig:RIC_222}.

It is clear that there are symmetries in the $(\gamma_1, \theta_1)$ space. In particular, $\gamma_1\rightarrow 4\pi-\gamma_1$ and $\theta_1\rightarrow 4\pi-\theta_1$, that effectively reduce the space to $0\le\gamma_1,\,\theta_1< 2\pi$, which can be easily understood by observing the structure of the trigonometric functions in (\ref{eq:EW_trans_red_boundaries}). The symmetry in $\gamma_1$ is equivalent to the global EW transformation $(\sigma^0\otimes\sigma^3)$ and the symmetry in $\theta_1$ simply corresponds to complex conjugation, both of which affect $\Phi(+\infty)$ in general, but not $\Phi(-\infty)$, similarly to the transformation of (\ref{eq:theta2_elim_trans}).

\begin{figure}
\centering
\vspace{0.5pt}
\includegraphics[width=\columnwidth]{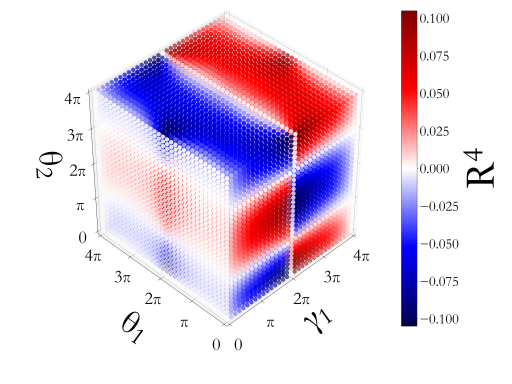}
\label{fig:EW_param_space_R4}
\vspace{5pt}
\includegraphics[width=\columnwidth]{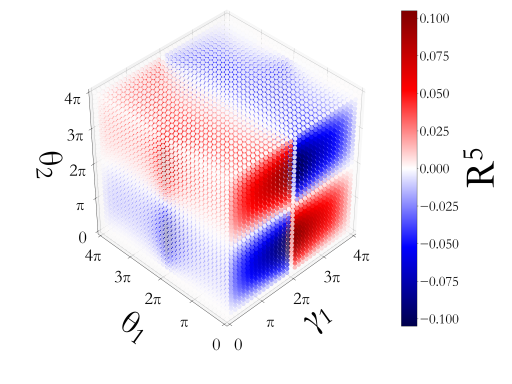}
\label{fig:EW_param_space_R5}
\caption{Values of $R^4$ and $R^5$ at the centre of fixed boundary domain wall solutions, as functions of $\gamma_1$ and $\theta_1$ in parameter set A for relatively EW rotated vacua. The rotations were varied in increments of $\pi/8$ in the range $[0, 4\pi]$, such that $32^3$ different solutions were computed.}
\label{fig:3D_param_spaces}
\end{figure}

We also present in Fig. \ref{fig:3D_param_spaces} the components of $R^4$ and $R^5$. While these extended components are invariant under an $SU(2)_L$ transformation they are manifestly variant under a simple $U(1)$ phase change, and as such $\theta_2$ also contributes to their variation. The variation of the energy under the three rotation parameters is not included as the energy was found to be invariant under $\theta_2$.

\begin{figure*}
    \centering
    \subfloat[]{\includegraphics[width=0.5\textwidth]{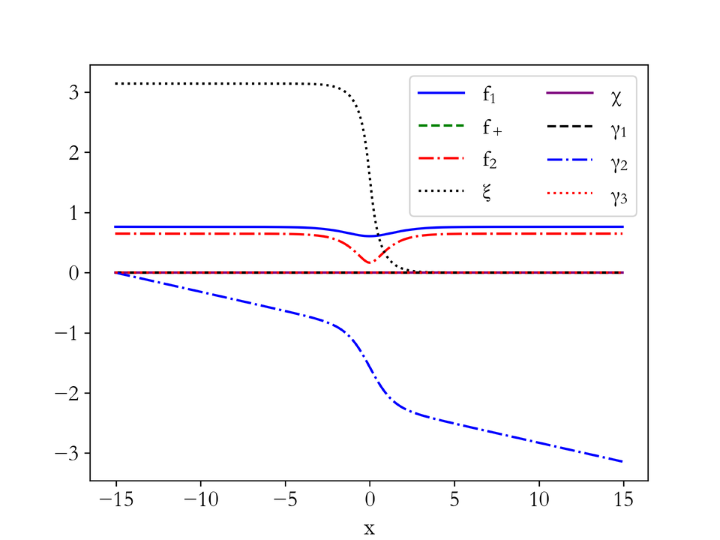}}
    \subfloat[]{\includegraphics[width=0.5\textwidth]{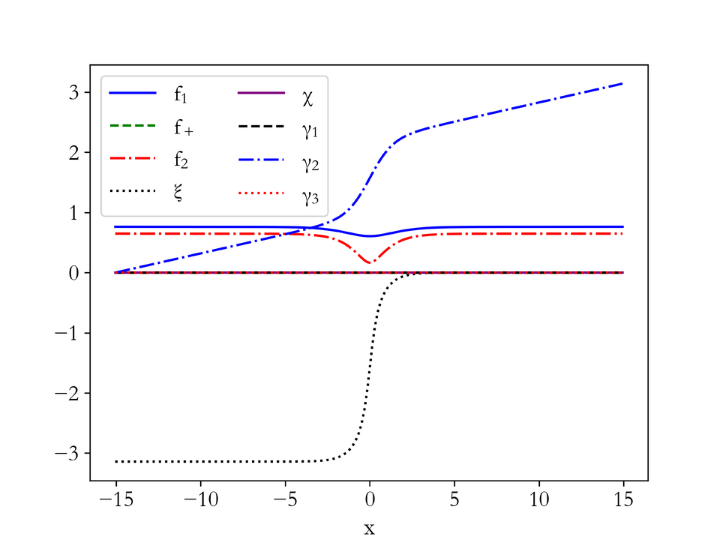}}
    \\\vspace{-10pt}
    \subfloat[]{\includegraphics[width=0.5\textwidth]{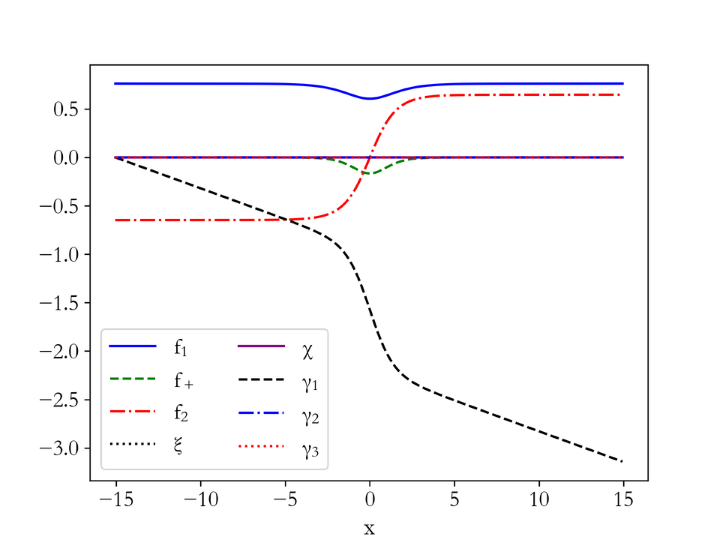}}
    \subfloat[]{\includegraphics[width=0.5\textwidth]{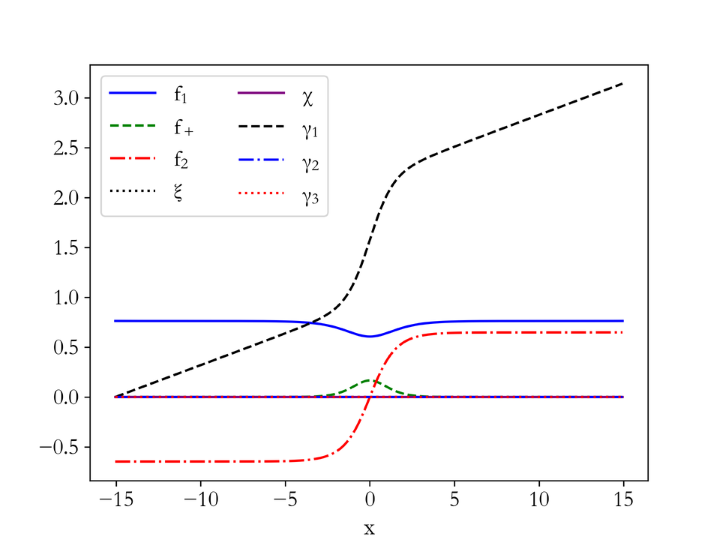}}
    \caption{Relatively EW rotated domain wall solutions where the solution exhibits either non-zero $R^2$ (top) and $R^4$ (bottom) in parameter set A. Solutions where the value at the centre of the wall is positive are depicted on the left and those of equal magnitude but negative on the right. Within the space of ($\gamma_1, \theta_1, \theta_2$) presented are solutions (a):($0,\pi,0)$, (b):($2\pi,\pi,0$), (c):($3\pi,0,0$), (d):($\pi,0,0$). In (a) and (b) the field components $f_+,\, \chi,\, \gamma_1,\, \gamma_3$ are identically zero for all $x$, in (c) and (d) $\xi,\, \chi,\, \gamma_2,\, \gamma_3$ are identically zero for all $x$.}
    \label{fig:EW_Analysis}
\end{figure*}

We can map the solutions that we have produced in the linear representation to the general using (\ref{eq:RA_to_gen}, \ref{eq:gen_to_lin}), allowing us to highlight the following features. Given our observation of non-zero values in the bi-linear components of $R^2$ and $R_\mu R^\mu$ (including the winding of $R^4$) within full dynamical simulations, we present in Fig.~\ref{fig:EW_Analysis} a selection of solutions designed to illustrate how these features manifest in the general representation.
These solutions are restricted to those where either $R^2$ is non-zero at the centre of the wall, for which the domain wall forms in $\xi$ and those where $R^\mu R_\mu$ is non-zero, for which the domain wall forms in $f_2$. We can immediately see from these solutions that for both a non-zero $R^2$ and $R^\mu R_\mu$ the two EW parameters of $\chi$ and $\gamma_3$ are identically zero and can be removed from the problem, which is unsurprising given the elimination of $\theta_2 = \chi + \gamma_3$. We observe that in the case where $R^\mu R_\mu \neq 0$ then $\sgn(f_+) = \sgn(\partial_x \gamma_1)$, which was also found to be the case in a set of solutions presented in ref.~\cite{Sassi:2023cqp}, and in the case of $R^2 \neq 0$ that $\sgn(\partial_x\xi) = \sgn(\partial_x\gamma_2)$, a previously unexplored observation of this type of solution. We note also that in each case, one could apply a global rotation to shift either $\gamma_1$ or $\gamma_2$ by a constant such that they are zero at the centre of the solution, which highlights that the relevant structure of each type of solution is the non-zero value of $\partial_x \gamma_i$ at the centre of the kink, ie. the interpolation of $\gamma_i$

The effects of $\gamma_1$ and $\gamma_2$ upon $f_+$ and $\xi$ respectively were the motivation for the form of our analytic reduction of the general field configuration in Sec.~\ref{sec:field_reduction}, as the structure of the field configuration is preserved in $\gamma_1$ and $\eta_1$ respectively.

\section{General Domain Wall Solutions with $\mathbb{Z}_2$ Symmetry}\label{sec:solution}
Before turning to the general minimum-energy domain wall solutions, it is instructive to first consider several special subclasses that arise within the reduced field configuration of Sec.~\ref{sec:field_reduction}. These subclasses correspond to distinct regions of parameter space where certain field components consistently vanish, leading to simplified ansätze with clear and characteristic structure. Indeed, some of these solutions coincide directly with those already identified in refs.~\cite{Battye2011VT, Battye2021SDW, Sassi:2023cqp}. Presenting these subclasses first is not merely pedagogical: our subsequent analysis reveals that the fully general solutions always reduce to one of these subclasses, depending on the region of parameter space under consideration. In this sense, the subclasses serve as natural building blocks, and their study both clarifies the organisation of the parameter space and is essential for interpreting the general solution landscape. All solutions presented in this section are naturally bounded. Only when we analyse the unrestricted reduced general field configuration do we obtain the minimum-energy solutions.

All solutions in this section were obtained using the \textit{Full Treatment} of Appendix~\ref{sec:numericals_kinks}. For broad parameter scans we used numerical settings of $n_x = 600,\, \Delta x = 0.05,\, \delta=10^{-5}$ to efficiently cover the parameter range. For individual example solutions we used $n_x = 3000,\, \Delta x = 0.01,\, \delta=10^{-7}$ to achieve higher accuracy in the profiles.

\subsection{$\gamma_1 = \eta_1 = \eta_2 = \eta_3 = 0$}\label{sec:standard_sol}
We first consider the case where all EW group parameters in (\ref{eq:gen_sol_matrix}) are set to zero. We find that this restricted ansatz does not generically yield what we have defined as the \textit{standard solution}, originally presented in ref.~\cite{Battye2011VT}. This earlier work found the field configuration to enforce $f_{+} \equiv \xi \equiv 0$, but we find this behaviour only arises for specific choices of model parameters. In fact, we find that specific regions of parameter space admit either superconducting or CP-violating behaviour, with $f_+ \neq0$ or $\xi\neq0$ respectively, subject to the mass hierarchy of $M_A$ and $M_{H^\pm}$. The dependence on the $M_A$ and $M_{H^\pm}$ mass ordering is explained in the semi-analytic categorisation of the general case in Sec.~\ref{sec:gen_sol}.

\begin{figure}
    \centering
    \subfloat[]{\includegraphics[width=\columnwidth]{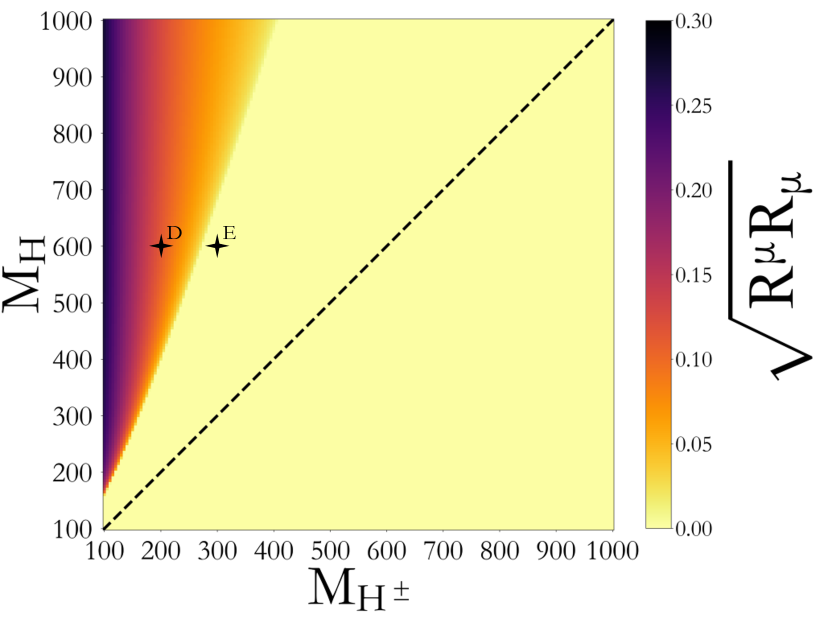}\label{fig:stan_sol_RmuRmu}}\\\vspace{-10pt}
    \subfloat[Parameter set D]{\includegraphics[width=\columnwidth]{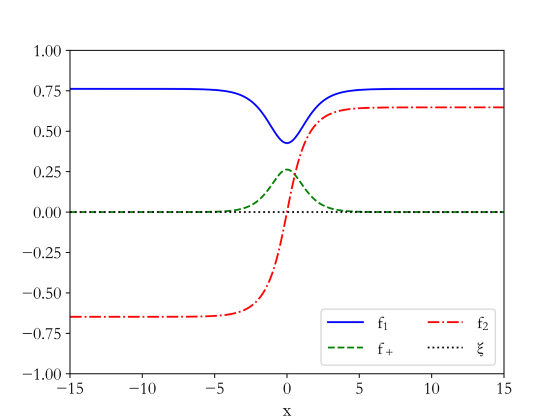}\label{fig:stan_sol_cond}}
    \caption{(a) $\sqrt{R^\mu R_\mu}$ at the centre of the domain wall for naturally bounded solutions of the field configuration (\ref{eq:gen_rep}) with the field restriction $\gamma_1 = \eta_1 = \eta_2 = \eta_3 = 0$, and parameter restriction $M_A > M_H^\pm$. Scalar masses are given in GeV, and we set $M_A$ to be $100 \text{ GeV}$ greater than $M_{H^\pm}$, however the magnitude of this mass difference does not affect the solutions. There is a clear dark-shaded region where there exist solutions with a stable condensate. The dashed line represents $M_H = M_{H^\pm}$. (b) Example solution for parameter set D, where a condensate forms in $f_+$.}
\end{figure}

When $M_{A} > M_{H^\pm}$ there is a region of the parameter space, shown in the dark-shaded area of Fig.~\ref{fig:stan_sol_RmuRmu}, in which the wall develops a stable condensate; an explicit example is provided in Fig.~\ref{fig:stan_sol_cond}. In this case $\xi \equiv 0$, so the potential no longer depends on the combination $\lambda_4 + |\lambda_5|$ and the solutions become independent of $M_A$.

\begin{figure}
    \centering
    \subfloat[]{\includegraphics[width=\columnwidth]{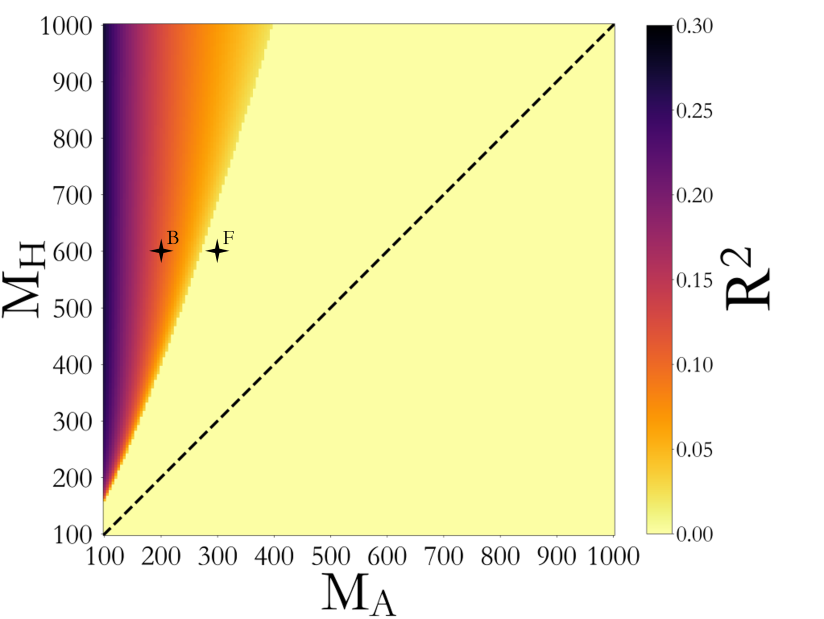}\label{fig:stan_sol_R2}}\\\vspace{-10pt}
    \subfloat[Parameter set B]{\includegraphics[width=\columnwidth]{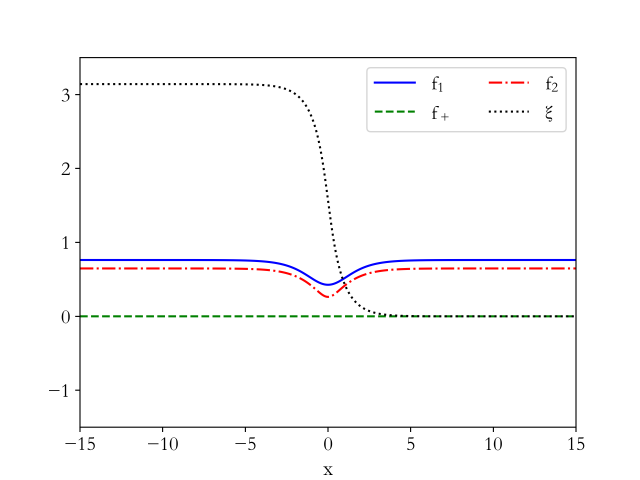}\label{fig:stan_sol_xi}}
    \caption{(a) $R^2$ at the centre of the domain wall for naturally bounded solutions of the field configuration (\ref{eq:gen_rep}) with the field restriction $\gamma_1 = \eta_1 = \eta_2 = \eta_3 = 0$, and parameter restriction $M_A < M_H^\pm$. Scalar masses are given in GeV, and we set $M_{H^\pm}$ to be $100 \text{ GeV}$ greater than $M_A$, however the magnitude of this mass difference does not affect the solutions. There is a clear dark-shaded region where there exist solutions with domain wall in $\xi$. The dashed line represents $M_H = M_A$. (b) Example solution for parameter set B, where the wall exists smoothly in $\xi$.}
\end{figure}

Conversely, when $M_{A} < M_{H^\pm}$ a region exists where the wall exists smoothly in $\xi$ rather than $f_{2}$. This generates a non-zero $R^2$ at the centre of the wall, giving rise to a qualitatively new type of solution. This parameter region is illustrated in Fig.~\ref{fig:stan_sol_R2} and an example solution form is given in Fig.~\ref{fig:stan_sol_xi}. Such $\xi$-wall configurations have been overlooked in previous studies, which have focused exclusively on walls in $f_2$. For this type of solution $f_{+} \equiv 0$, the potential then depends directly on the combination of $\lambda_3 + \lambda_4$, eliminating the dependence on $M_{H^\pm}$.


\begin{figure}
    \includegraphics[width=\columnwidth]{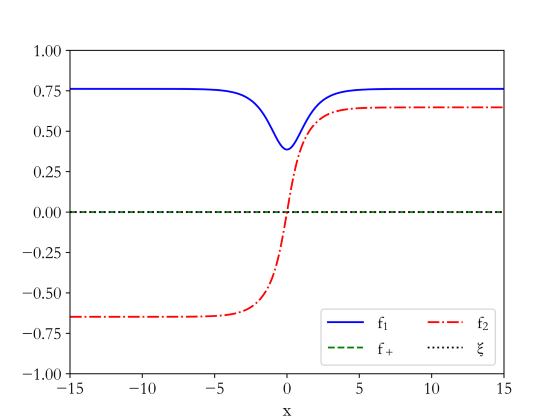}
    \caption{Naturally bounded domain wall solution in parameter sets E and F for the configuration (\ref{eq:gen_rep}) with the field restriction $\gamma_1 = \eta_1 = \eta_2 = \eta_3 = 0$. This type of solution is fully described by the field components $f_1$ and $f_2$ only and is identified as the \textit{standard solution} form.}
    \label{fig:stan_sol_non_cond}
\end{figure}

Outside of these two regions, the solutions take the form shown in Fig.~\ref{fig:stan_sol_non_cond}, which we identify as the \textit{standard solution}. This subclass is described entirely by $f_{1}$ and $f_{2}$, with $f_{+} \equiv \xi \equiv 0$, which consequently makes the solutions independent of both $M_A$ and $M_{H^\pm}$. The appearance of the condensates or $\xi$-walls in this restricted configuration can be understood as reductions to the potential energy when $M_{H^\pm}$ or $M_A$ is sufficiently small compared to $M_H$ respectively, with the precise boundary depending on $\tan\beta$. The general conditions are developed in the semi-analytic discussion of Sec.~\ref{sec:gen_sol}. Note that the regions of $R^\mu R_\mu \neq0$ and $R^2\neq 0$ in the two different mass orderings of this restricted ansatz coincide exactly in extent.

We do not distinguish the superconducting ($f_+\neq0$) or CP-violating ($\xi\neq0$) solutions within this ansatz as distinct subclasses of solution. As shown in Secs.~\ref{sec:gamma1} and ~\ref{sec:eta1}, whenever the configurations shown in Figs.~\ref{fig:stan_sol_cond} and ~\ref{fig:stan_sol_xi} are possible under this restricted ansatz, lower-energy solutions arise once $\gamma_1$ or $\eta_1$ are allowed to vary respectively. In other words, these configurations are never minimum-energy solutions when $\gamma_1$ or $\eta_1$ are unrestricted. We have included the above discussion of condensates and $\xi$ walls in this restricted ansatz as a precursor to the following subclasses and to complete the delineation of this ansatz.

The central point is to distinguish the \textit{standard solution} as a distinct simple subclass of solution, characterised by the form in Fig.~\ref{fig:stan_sol_non_cond} and dependent solely on the general representation fields $f_1$ and $f_2$ as previously defined, with $f_+ \equiv \xi \equiv 0$.

\subsection{$\gamma_1 \neq 0$, $\xi = \eta_1 = \eta_2 = \eta_3 = 0$}\label{sec:gamma1}
\begin{figure}
    \centering
    \subfloat[]{\includegraphics[width=\columnwidth]{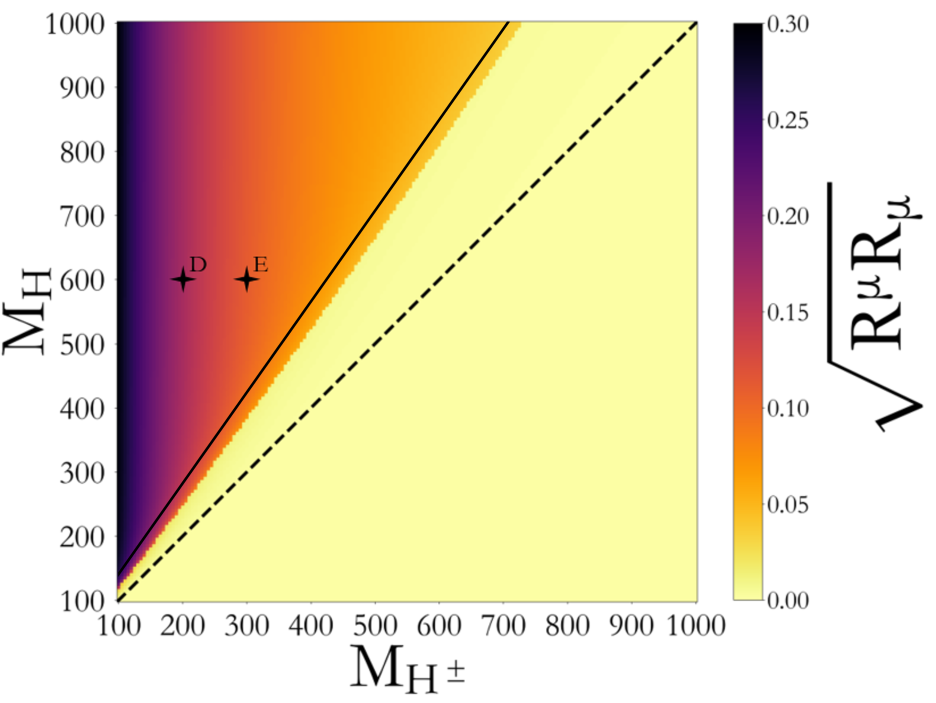}\label{fig:gam1_mass_scan}}\\\vspace{-10pt}
    \subfloat[Parameter set E]{\includegraphics[width=\columnwidth]{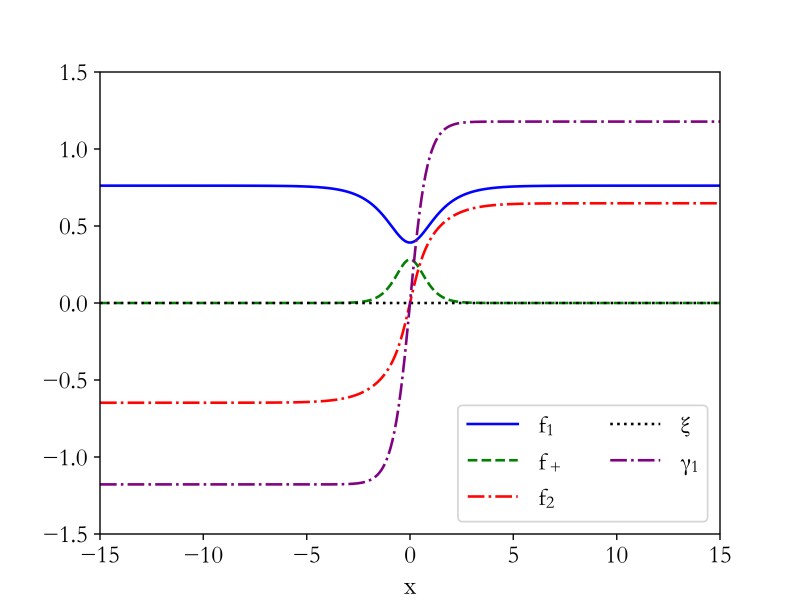}\label{fig:gam_1_sol}}
    \caption{(a) $\sqrt{R^\mu R_\mu}$ at the centre of the domain wall for naturally bounded solutions of the field configuration (\ref{eq:gen_rep}) with the field restriction $\eta_1 = \eta_2 = \eta_3 = 0$, as functions of the two dependent mass parameters $M_H$ and $M_{H^\pm}$ in GeV. Note there is a clear region where there exist solutions with a stable superconducting condensate and in all cases $\xi \equiv 0$. The solid line represents a semi-analytic prediction of the boundary using (\ref{eq:sol_region_eval}) and (\ref{eq:f1_robust_prediction}), while the dashed line represents $M_H= M_{H^\pm}$. (b) Example solution for parameter set E, which is identified as the \textit{superconducting solution} form.}
\end{figure}

In ref.~\cite{Sassi:2023cqp} it was shown that allowing only the parameter $\gamma_1$ of (\ref{eq:EW_matrx}) to vary yields a naturally bounded solution with a stable condensate at the centre of the wall. Their analysis, performed for a single parameter set ($M_H = 800 \text{ GeV},\, M_A = 500 \text{ GeV},\, M_{H^\pm} = 400 \text{ GeV},\, \tan\beta=0.85$), identified a solution similar to the type we display in Fig.~\ref{fig:gam_1_sol}, which is for parameter set E.

Examining the energy functional confirms that a non-zero $f_+$ at the centre of the wall lowers the energy further should $\gamma_1$ interpolate across the wall, with $\sgn(f_+)=\sgn(\partial_xf_2\partial_x \gamma_1)$. This follows directly from the energy density term for this subclass,
\begin{equation}
    -\half f_+\partial_x f_2 \partial_x \gamma_1\,,
\end{equation}
in agreement with the analysis of ref.~\cite{Sassi:2023cqp}. Furthermore, as also noted (but not explicitly demonstrated) in ref.~\cite{Sassi:2023cqp}, the values that $\gamma_1$ takes at the boundaries, and thus the amplitude of the condensate $f_+$, are directly related to the masses of the physical scalars, a result which we confirm.

We restrict $\xi \equiv 0$ here such that the domain wall is explicitly in $f_2$, the solutions can admit a condensate and are again independent of $M_A$. This allows us to perform a parameter scan of $M_H$ and $M_{H^\pm}$ which is shown in Fig~\ref{fig:gam1_mass_scan}. We observe an enhanced region of the parameter space where a stable condensate exists upon the wall, accompanied by an interpolating profile for $\gamma_1$. Outside of this region we find the field configuration to relax to the previously determined standard solution, with $f_+$ and $\gamma_1$ vanishing. We observe that the effect of allowing $\gamma_1$ to vary can be dramatic; the solutions in Fig.~\ref{fig:stan_sol_non_cond} and Fig.~\ref{fig:gam_1_sol} share identical scalar masses, yet only the latter supports a condensate. We also note that equivalent solutions exist with opposite condensate sign, realised when $\gamma_1$ interpolates in the opposite direction, as already pointed out in ref.~\cite{Sassi:2023cqp}. These equivalent solutions are simply related by a global rotation.

The mass dependence of $\sqrt{R^\mu R_\mu}$ at the centre of the wall closely mirrors the behaviour seen in ref.~\cite{Battye:2024iec}, where condensation onto the core of the string occurs when $M_H \gtrsim M_{H^\pm}$. We shall return to and quantify this correspondence later in Sec.~\ref{sec:gen_sol} where we provide a semi-analytic explanation of the parameter dependence of the solution subclasses.

We now propose the form of the solution in Fig.~\ref{fig:gam_1_sol} as our second subclass of solution, the \textit{superconducting solution}, characterized by a stable condensate on the wall and an interpolating $\gamma_1$ profile with local breaking of the electromagnetic $U(1)$ symmetry.

\subsection{$\eta_1 \neq 0$, $f_+ = \gamma_1 = \eta_2 = \eta_3 = 0$}\label{sec:eta1}
The findings of Sec.~\ref{sec:EW_Rel_Trans} suggest that an interpolating profile in $\eta_1$ should accompany a wall in $\xi$. From the energy functional it clearly follows that interpolating $\xi$ and $\eta_1$ profiles lower the energy further whenever $\sgn(\partial_x\xi) = \sgn(\partial_x\eta_1)$, due to the term in the energy density,
\begin{equation}
    -\half f_2^2 \partial_x \xi \partial_x \eta_1\,.
\end{equation}
As suggested in Sec.~\ref{sec:standard_sol} and to be shown explicitly in Sec.~\ref{sec:gen_sol}, the presence or absence of a condensate, and likewise the existence of a wall in $\xi$, is controlled by the hierarchy of $M_A$ and $M_{H^\pm}$. Here we restrict $f_+\equiv 0$ such that we are explicitly in the $\xi$-wall regime and the solutions are independent of $M_{H^\pm}$.

\begin{figure}
    \centering
    \subfloat[]{\includegraphics[width=\columnwidth]{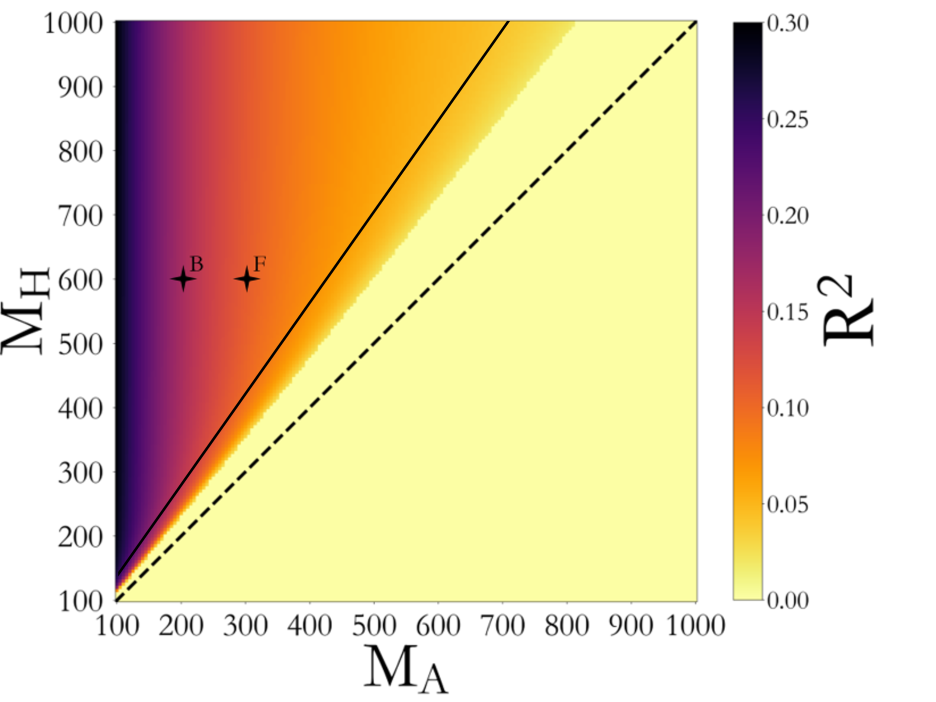}\label{fig:eta1_mass_scan}}\\\vspace{-10pt}
    \subfloat[Parameter set F]{\includegraphics[width=\columnwidth]{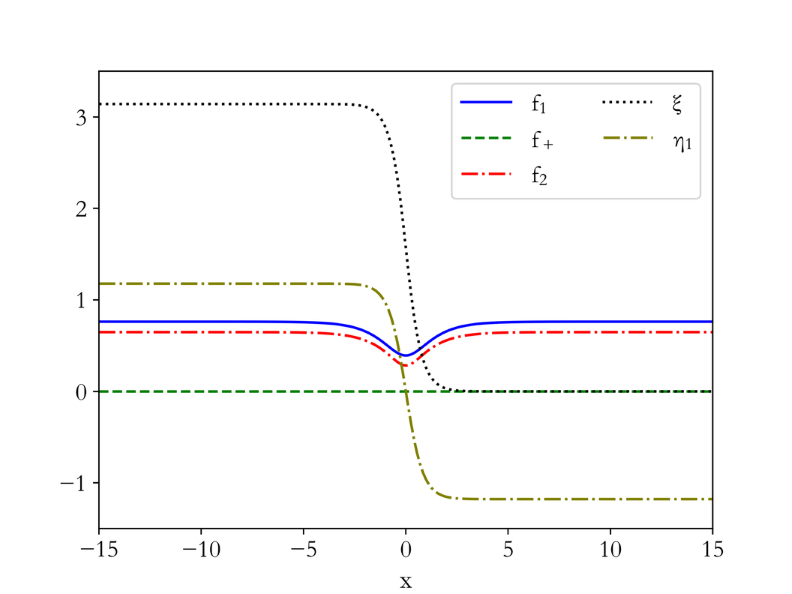}\label{fig:eta_1_sol}}
    \caption{(a) $R^2$ at the centre of the domain wall for naturally bounded solutions of the field configuration (\ref{eq:gen_rep}) with the restriction $f_+ = \gamma_1 = \eta_2 = \eta_3 = 0$, as functions of the two dependent mass parameters $M_H$ and $M_A$ in GeV. Note there is a clear region where there exist solutions with a domain wall in $\xi$ as opposed to $f_2$. The solid line represents a semi-analytic prediction of the boundary using (\ref{eq:sol_region_eval}) and (\ref{eq:f1_robust_prediction}), while the dashed line represents $M_H= M_A$. (b) Example solution for parameter set F, which is identified as the \textit{CP-violating solution} form.}
\end{figure}

In Fig.~\ref{fig:eta1_mass_scan} we show a mass parameter scan for this ansatz, with an example solution in Fig.~\ref{fig:eta_1_sol}. This, akin to the variation of $\gamma_1$, reveals an enhanced region where $\xi$-wall solutions exist. Outside of this region, the lowest-energy configuration again relaxes to the standard solution. We likewise see that the effect of allowing $\eta_1$ to vary can be dramatic; the solutions in Fig.~\ref{fig:stan_sol_non_cond} and Fig.~\ref{fig:eta_1_sol} share identical scalar masses, yet in the former the domain wall is in $f_2$ and the in latter in $\xi$. We also note that equivalent solutions exist with opposite sign of $R^2$, realised when $\eta_1$ interpolates in the opposite direction and $\xi$ interpolates from $-\pi$ to zero, passing through $-\pi/2$ at the centre of the wall. These equivalent solutions are simply related by a CP1 transformation of $\Phi \rightarrow \Phi^*$.

We therefore propose the type of solution seen in Fig.~\ref{fig:eta_1_sol} as our third subclass, the \textit{CP-violating solution}, characterised by the domain wall residing in $\xi$ with an interpolating $\eta_1$ profile, causing a local breaking of CP symmetry, while the vacua remain CP symmetric. As with the superconducting case of Sec.~\ref{sec:gamma1}, we will delineate the precise boundary of this solution space in Sec.~\ref{sec:gen_sol}.

\subsection{General Solutions}\label{sec:gen_sol}
We now consider the minimum-energy domain wall solutions obtained from the unrestricted field configuration, consistently reduced to six independent fields as outlined in Sec.~\ref{sec:field_reduction}. Solving the full equations of motion allows us to partition the parameter space into four distinct subclasses: standard (Sec.~\ref{sec:standard_sol}), superconducting (Sec.~\ref{sec:gamma1}), CP-violating (Sec.~\ref{sec:eta1}), and simultaneously superconducting \& CP-violating solutions.

The delineation of the parameter space can be understood first by categorizing the space into two distinct regions, those that contain a wall in $\xi$ and those that contain a wall in $f_2$. Neglecting gradient energy and considering only the potential (\ref{eq:2HDM_Potential_2}) we first consider under what conditions there will be a reduction to the energy for a $\xi$ profile interpolating between $0$ and $\pi$, naturally passing through $\pi/2$ at the centre of the domain wall. Let us consider the term proportional to
\begin{eqnarray}
    \lambda_4 - |\lambda_5|\cos2\xi = \frac{2}{v_{\rm SM}}\left[M_A^2\sin^2\xi - M^2_{H^\pm}\right]\,,
\end{eqnarray}
which is the only part of the potential depending on $M_A$ and $\xi$. At the centre of the domain wall where $\xi = \pi/2$ this takes a form $\propto (M^2_A - M^2_{H^\pm})$, suggesting that for a CP-violating wall to reduce the energy requires $M_A < M_{H^\pm}$. 

\begin{figure*}
    \centering
    \includegraphics[width=\textwidth]{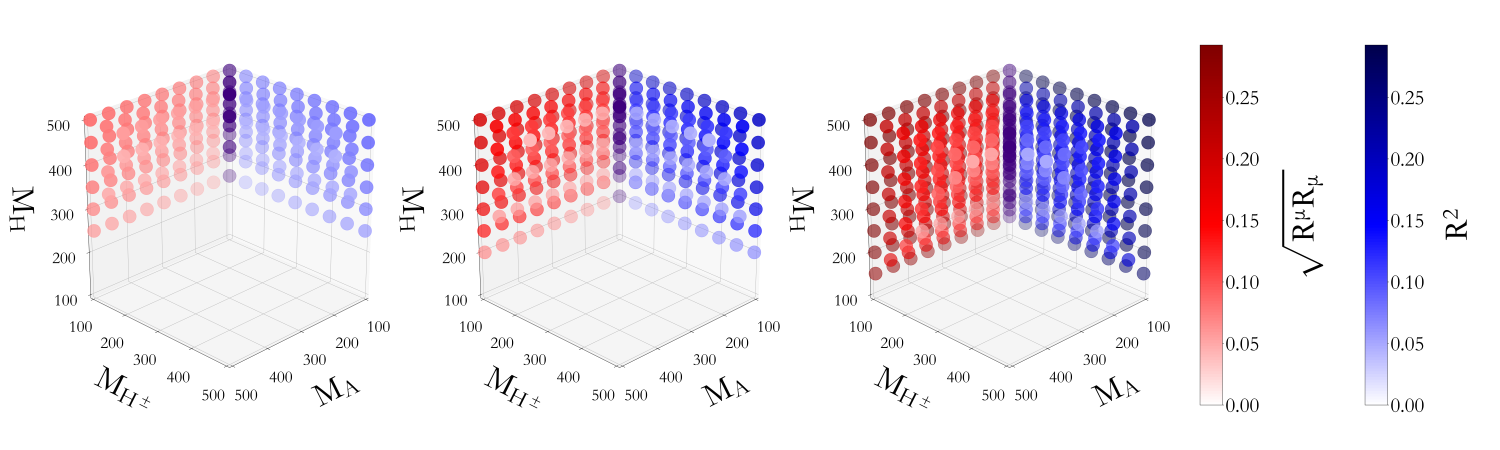}
\caption{$\sqrt{R^\mu R_\mu}$ and $R^2$ at the centre of minimum-energy domain wall solutions, for varying scalar masses, $M_H$, $M_A$ and $M_{H^\pm}$, where $\tan\beta = 0.10$ (left), $\tan\beta = 0.25$ (centre), $\tan\beta = 0.85$ (right). Masses with no data point represent standard solutions. Note the distinct regions where a superconducting and/or CP-violating solution exist. For an increasing $\tan\beta$ we observe an increasing region of non-standard solutions, in addition to increasing amplitudes of $\sqrt{R^\mu R_\mu}$ and $R^2$ at the centre of the wall.}
\label{fig:tan_beta_var}
\end{figure*}

We have confirmed this condition to hold by performing a parameter scan of $M_H,\, M_A,\, M_{H^\pm}$ for three fixed values of $\tan\beta$, for the unrestricted configuration. This is shown in Fig.~\ref{fig:tan_beta_var}, where we see that, if the solution is non-standard, when $M_A \leq M_{H^\pm}$ a domain wall forms in $\xi$, whereas if $M_A \geq M_{H^\pm}$ a stable condensate exists on the wall, with $M_A = M_{H^\pm}$ yielding solutions where both occur simultaneously.

Secondly, it is clear that the boundary between standard and non-standard (superconducting and/or CP-violating) solutions is dependent on $\tan\beta$ and the ratio of $M_H/M_X$ where $X = A,\, H^\pm$. This can be understood by considering the effective masses of $f_+$ and $f_2$ about the standard solution at the centre of the wall, where $f_+ = f_2 = \xi = 0$. Neglecting gradient energies, the sign of these effective masses determines whether it is energetically favourable for the standard solution to develop a non-zero value in either $f_+$ or $f_2$.

In the mass regime $M_{H^\pm} \leq M_A$ the relevant fluctuation is in $f_+$. Expanding the potential about the standard solution at the centre of the wall gives the effective mass 
\begin{eqnarray}
    M^2_+(0) & = & -\half\mu_2^2 + \quart\lambda_3f_1^2(0)\cr 
    & = & \quart M_h^2\left[\left(1-\hat M_H^2 + 2\hat M_{H^\pm}^2\right)f_1^2(0) - 1\right],
    \label{eq:M+_exp}
\end{eqnarray}
where $\hat M_H = M_H/M_h$ and $\hat M_{H^\pm} = M_{H^\pm}/M_h$. A negative $M_+^2(0)$ signifies an instability in the $f_+$ direction, leading to condensation and the superconducting solutions discussed in Sec.~\ref{sec:gamma1}. 

Conversely, in the opposite mass regime, $M_{H^\pm} \geq M_A$, the relevant fluctuation is in $f_2$. Expanding about the standard solution at the centre of the wall yields
\begin{eqnarray}
        M^2_2(0) & = & -\half\mu_2^2 + \quart\Big(\lambda_3 + \lambda_4 + |\lambda_5|\Big)f_1^2(0)\cr
        & = & \quart M_h^2\left[\left(1-\hat M_H^2 + 2\hat M_A^2\right)f_1^2(0) - 1\right],
        \label{eq:M2_exp}
\end{eqnarray}
where, as previously, $\hat M_A = M_A/M_h$. In making this identification we implicitly assume that any non-zero $f_2$ is accompanied by $\xi=\pi/2$ at the wall centre. Recalling our distinction in Sec.~\ref{sec:motivation}, it is clear that this assumption is true such that the $\mathbb{Z}_2$ symmetry is respected in each coordinate basis of the wall. This is precisely the behaviour realised in the CP-violating configurations of Sec.~\ref{sec:eta1}. A negative $M_2^2(0)$ signifies an instability of the standard solution in the $f_2$ direction, leading to CP-violating configurations.

We see that these two effective mass expressions $M_+^2(0)$ and $M^2_2(0)$ exhibit an exact symmetry under the exchange of $M_{H^\pm} \leftrightarrow M_A$. This symmetry naturally unifies the classification of the superconducting and CP-violating regimes, allowing them to be treated simultaneously within the same framework, given our approximation of neglecting gradient energies. To fully categorise the regions of mass space, we must therefore evaluate where either $M_+^2$ or $M_2^2$, will be zero in terms of $M_H, M_X$ and $\tan\beta$.

We can evaluate the mass dependence of $f_1(0)$ using the reduced standard solution ansatz of
\begin{equation}
    \Phi = \frac{v_{\rm SM}}{\sqrt{2}}\begin{pmatrix} 0 \\ f_1 \\ 0 \\ f_2 \end{pmatrix}\,,
    \label{eq:most_reduced_fields}
\end{equation}
and rearranging (\ref{eq:M+_exp}, \ref{eq:M2_exp}) we find that one would therefore expect a standard solution when
\begin{equation}
    \hat M_X^2 \gtrsim \half\left(\hat M_H^2 + f^{-2}_1(0) - 1 \right)\,,
    \label{eq:sol_region_eval}
\end{equation}
where as before $X = A,\, H^\pm$ for the appropriate regime.

We can obtain a rough estimate of the critical masses by assuming $\partial_x^2f_1=0$ at the centre, in which case a quick analysis of the equation of motion for $f_1$ reveals that
\begin{equation}
    \tilde f_1(0) = \frac{1}{1 + \hat M_H^2\tan^2\beta}\,,
    \label{eq:sol_region_approx}
\end{equation}
where we define $\tilde f_1(0)$ to be an approximate expression for $f_1(0)$. This approximation holds exactly if $\hat M_H = 1$, where $f_1$ is found to be a constant and as such $\partial_x^2 f_1 = 0$. Away from $\hat M_H = 1$ however, $f_1$ is found to develop a local extremum at the wall, as can be seen in all our example solutions. We evaluated $f_1(0)$ across the parameter intervals $\hat M_H = [1, 8]$, $\tan\beta = [0.25, 8]$ and found this simple approximation to differ by up to approximately $70\%$ on this interval, with it expected to worsen for higher values of $\hat M_H$. However, the benefit of this simple analytic estimate is the ability to make the first approximation that we would expect a standard solution if
\begin{equation}
     \hat M_{{H^\pm},\text{ } A}^2 \gtrsim  \half \hat M_H^2 \left(1 + \tan^2\beta\right),
\end{equation}
and a superconducting and/or CP-violating solution otherwise, given the appropriate mass ordering regime.

While this simple analytic evaluation provides a good first approximation we have found that a set of minimal correctional terms may be introduced to significantly improve the evaluation of $f_1(0)$ on the interval we evaluated. Introducing these terms provides us with the more robust expression to be substituted into (\ref{eq:sol_region_eval}),
\begin{eqnarray}
\tilde f_1(0) & = &
\frac{1}{1 + \hat M_H^2 \tan^2\beta}
\Biggl[ 1 + (\hat M_H - 1)\Bigl(
   c_1\,\hat M_H \cr
& + &c_2\,\hat M_H^2 
 + c_3\,\tan\beta
 + c_4\,\tan^2\beta
\Bigr)\Biggr]\,,
\label{eq:f1_robust_prediction}
\end{eqnarray}
with
\begin{eqnarray}
    c_1 & = & 0.1158\,, \quad c_2 = -0.0090\,, \cr 
    c_3 & = &0.0824\,, \quad c_4 = 0.0078\,,
\end{eqnarray}
where the correction terms were obtained from a minimax (Chebyshev) fit to the numerical data across the analysed intervals, anchored to the exact expression at $\hat M_H=1$. 
These numerical corrections improve the prediction of $f_1(0)$ such that it only differs by up to $7\%$ over our tested intervals, which given the neglecting of gradient energy in our approximation, provides an adequate prediction. This can be seen in Figs.~\ref{fig:gam1_mass_scan} and~\ref{fig:eta1_mass_scan}, where the solid lines represent the prediction using (\ref{eq:sol_region_eval}) and (\ref{eq:f1_robust_prediction}). This may be a less concise prediction compared to our first approximation, but it can be used for a more robust prediction if required, away from $\hat M_H = 1$. It must be stressed that this is still only an approximate predictive tool and will not be valid for any given combination of $\hat M_H$ and $\tan\beta$, specifically those outside our fitted intervals. Alternatively, one could simply evaluate the domain wall solution using the restricted standard solution ansatz for any given parameter set and identify using (\ref{eq:sol_region_eval}) the approximate form of the minimum-energy field configuration. This would be the most accurate approach but requires some numerical effort for each combination of $\hat M_H$ and $\tan\beta$.

In Appendix~\ref{sec:vanishing_fields} we have included a brief discussion on the nature of the field components present in the general field configuration which identically vanish in each of the solution regimes. It transpires that in each solution regime the fields restricted to be zero in each of the subclasses identically vanish, greatly simplifying the parameter space. Furthermore in Appendix~\ref{sec:gen_sols_set} we have included a selection of minimum-energy solutions for varying parameter sets to further demonstrate our findings.

Given our classification of the minimum-energy solutions we now refer back to our RIC simulations where it can be seen that in Fig.~\ref{fig:RIC_623} the walls are dominated by a non-zero $R^2$, reflecting the CP-violating minimum-energy solution of parameter set B, in Fig.~\ref{fig:RIC_632} the walls are dominated by non-zero $R^\mu R_\mu$ reflecting the corresponding superconducting solution of parameter set D, while the simulation of parameter set C in Fig.~\ref{fig:RIC_633} presents both non-standard features inline with its minimum-energy solution, found in Fig.~\ref{fig:gen_sol_fields_633}. This shows the success of our classification, with full dynamical simulations agreeing completely with our minimum-energy solutions while also showing evidence of relative EW rotations, as previously discussed.

\section{Current-Carrying Solutions}\label{sec:CC}
We now turn our attention to the case of current-carrying solutions. The earliest realization that topological defects such as cosmic strings can stably support currents goes back to Witten’s seminal work on superconducting strings \cite{Witten:1984eb}. Subsequently Davis and Shellard \cite{DAVIS1989209} provided a detailed analysis of vortex superconductivity and characterized stable loop configurations (Vortons) supported by these currents. Here we use the term \textit{superconducting} to indicate that the defect supports a condensate, as occurs in our superconducting subclass of domain walls where the $U(1)_{\rm EM}$ symmetry is broken locally on the wall. By contrast, we use \textit{current-carrying} to refer specifically to solutions in which this condensate sustains a propagating current along the defect.

Building on our findings of Sec.~\ref{sec:gen_sol}, and in agreement with refs.~\cite{Battye:2024dvw, Battye:2024iec}, we propose the following superconducting domain wall ansatz,
\begin{equation}
\Phi = \frac{v_{\rm SM}}{\sqrt{2}}\left[\sigma^0 \otimes \begin{pmatrix}
        \cos\half\gamma_1 & \sin\half\gamma_1 \\ -\sin\half\gamma_1 & \cos\half\gamma_1
    \end{pmatrix}\right] \begin{pmatrix}0 \\ f_1 \\ f_+ \\ f_2 \end{pmatrix}\,,
\label{eq:super_con_DW}
\end{equation}
subject to the parameter restrictions of $\hat M_{H^\pm} < \hat M_A$ and $\hat M_{H^\pm}^2 \lesssim \half\left(\hat M_H^2 +f^{-2}_1(0) - 1 \right)$, whereby we expect a superconducting condensate to form with $f_+ \neq 0$ at the centre of the wall.

To produce a current carrying field configuration we perform a space-time dependent transformation on the superconducting ansatz, using the degree of freedom which is unbroken in the vacuum, acting on (\ref{eq:super_con_DW}) with
\begin{equation}
    e^{\frac{1}{2} i \psi_\mu x^\mu} \left[ \sigma^0 \otimes e^{\frac{1}{2} i \psi_\nu x^\nu \sigma^3} \right]\,,
    \label{eq:cc_guage_trans}
\end{equation} 
where $\psi_\mu \psi^\mu = \omega^2 - k^2 \equiv \kappa$, such that we obtain, for a domain wall lying along the y-axis, the current-carrying ansatz
\begin{eqnarray}
\Phi & = & \frac{v_{\rm SM}}{\sqrt{2}}\begin{pmatrix}f_1 \sin\half\gamma_1e^{i\left(\omega t + ky\right)}  \\ f_1 \cos\half\gamma_1\\ \left(f_+\cos\half\gamma_1 + f_2 \sin\half\gamma_1\right)e^{i\left(\omega t + ky\right)} \\ -f_+ \sin\half\gamma_1 + f_2 \cos\half\gamma_1\end{pmatrix}\nonumber \\
	 & = & \frac{v_{\rm SM}}{\sqrt{2}}\begin{pmatrix}g_1e^{i\left(\omega t + ky\right)}  \\ g_2\\ g_3e^{i\left(\omega t + ky\right)} \\ g_4\end{pmatrix}\,.
\label{eq:CC_DW}
\end{eqnarray}
Here we have introduced $g_i$ functions to replace $f_1,\, f_+,\, f_2,\,\gamma_1$ in order to simplify numeric calculations. 

This ansatz yields the following Lagrangian and energy densities,
\begin{eqnarray}
   \mathcal{L} & = &-\half (\partial_xg_1)^2  -\half (\partial_xg_2)^2 - \half (\partial_xg_3)^2 - \half (\partial_xg_4)^2 \nonumber \\
    & + &\half \kappa \left(g_1^2 + g_3^2\right) - V\,,
    \label{eq:CC_lag}
\end{eqnarray}
\begin{eqnarray}
   \mathcal{E} & = &\half (\partial_xg_1)^2  + \half (\partial_xg_2)^2 + \half (\partial_xg_3)^2 + \half (\partial_xg_4)^2 \nonumber \\
    & + &\half (\omega^2 +k^2) \left(g_1^2 + g_3^2\right) + V\,,
    \label{eq:CC_energy}
\end{eqnarray}
where we have that $g_i \equiv g_i(x)$. This construction is directly analogous to the method of ref.~\cite{Battye:2024dvw} for superconducting strings, adapted here to the wall case. This introduces the further parameter of $\kappa$ to fix in the equations of motion ($\kappa$ is often called $\chi$ in the literature on current-carrying strings, however we use $\kappa$ to distinguish from the EW group parameter $\chi$), which plays the role of an effective mass term for the current-carrying components $g_1, g_3$, controlling the magnitude and regime of the solution. In the literature on current-carrying strings, solutions are classified into three regimes: chiral ($\kappa=0$), magnetic ($\kappa<0$), or electric ($\kappa>0$).

A subtlety arises for this ansatz because the transformation (\ref{eq:cc_guage_trans}) used to generate the current–carrying ansatz is not limited to only adding a phase to the condensate components, but also couples to the EW component of $\gamma_1$. Whether $\gamma_1$ relaxes to zero in the vacuum therefore depends on the sign of $\kappa$. In the magnetic regime, the effective mass term $\half \kappa(g_1^2 + g_3^2)$ has been found to suppress the current-carrying components away from the wall, which acts to drive $\gamma_1 \to 0$ in the vacuum, yielding consistent solutions. We have found however that in the chiral and electric regimes that this suppression of the current-carrying components does not occur, and $\gamma_1$ remains non-zero in the vacuum. Thus, within this ansatz, we have found that only magnetic current-carrying configurations can be constructed, without inducing non-zero vacuum energy. We cannot, however, exclude the possibility that there exist parameter sets in which chiral or electric solutions are consistent; we have simply not found any examples in our explorations.

We suggest that the limitation of this ansatz could be resolved in the gauged version of the theory, as one may choose to work in a gauge where $\gamma_1 = 0$ globally without loss of generality, however as this work is concerned only with the global theory we limit our discussion now to the magnetic regime.

\begin{figure*}
      \centering
      \subfloat[$\kappa = -0.1$]{\includegraphics[width=0.33\textwidth]{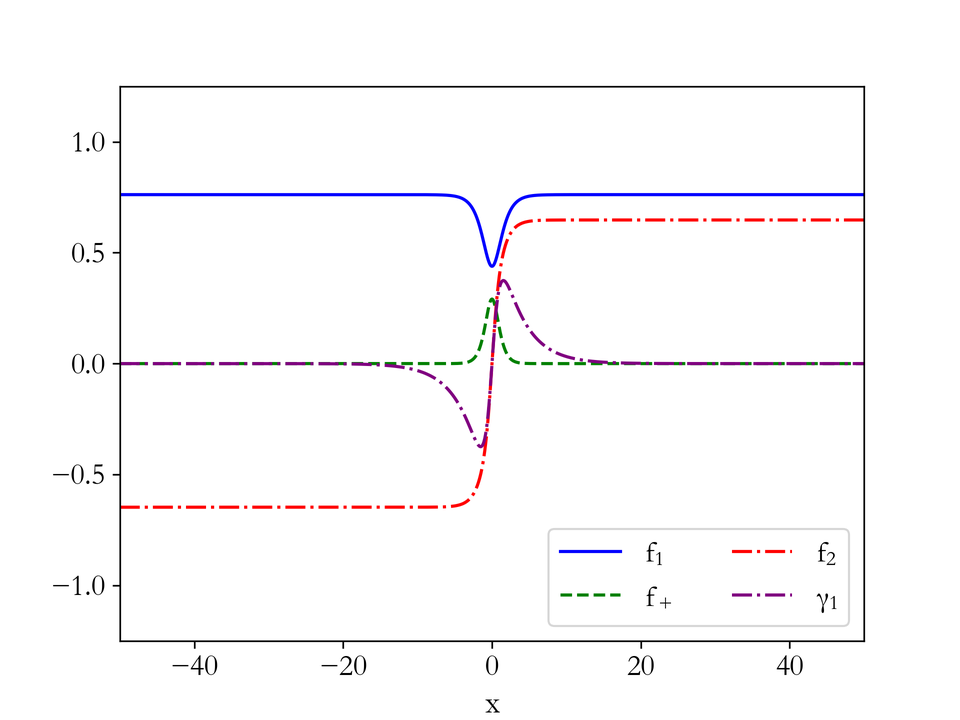}}
	\subfloat[$\kappa = -0.01$]{\includegraphics[width=0.33\textwidth]{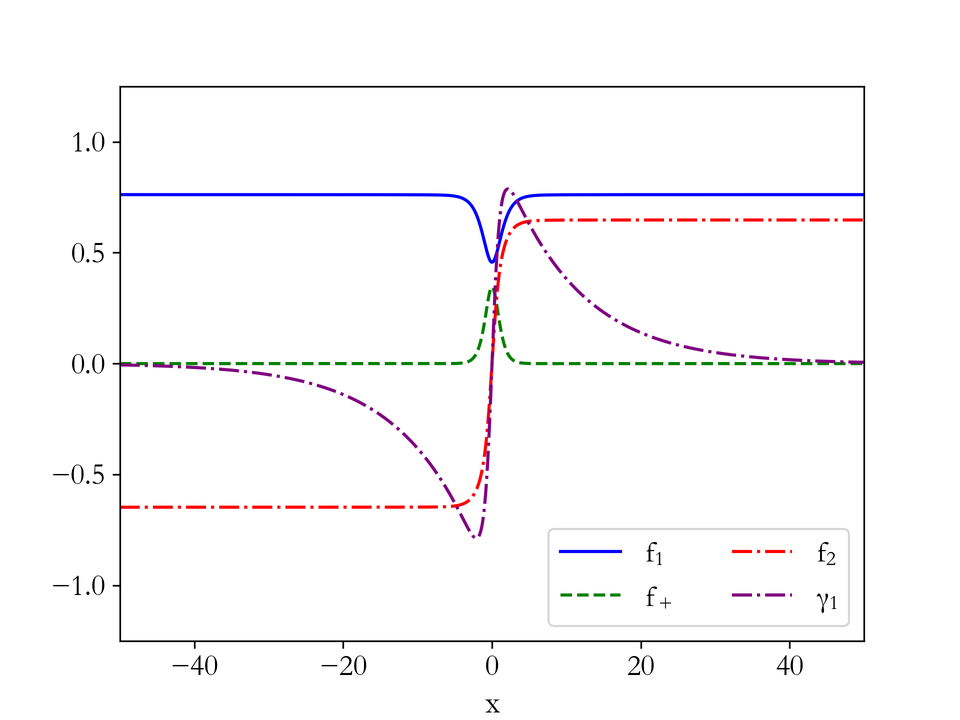}}
      \subfloat[$\kappa = -0.001$]{\includegraphics[width=0.33\textwidth]{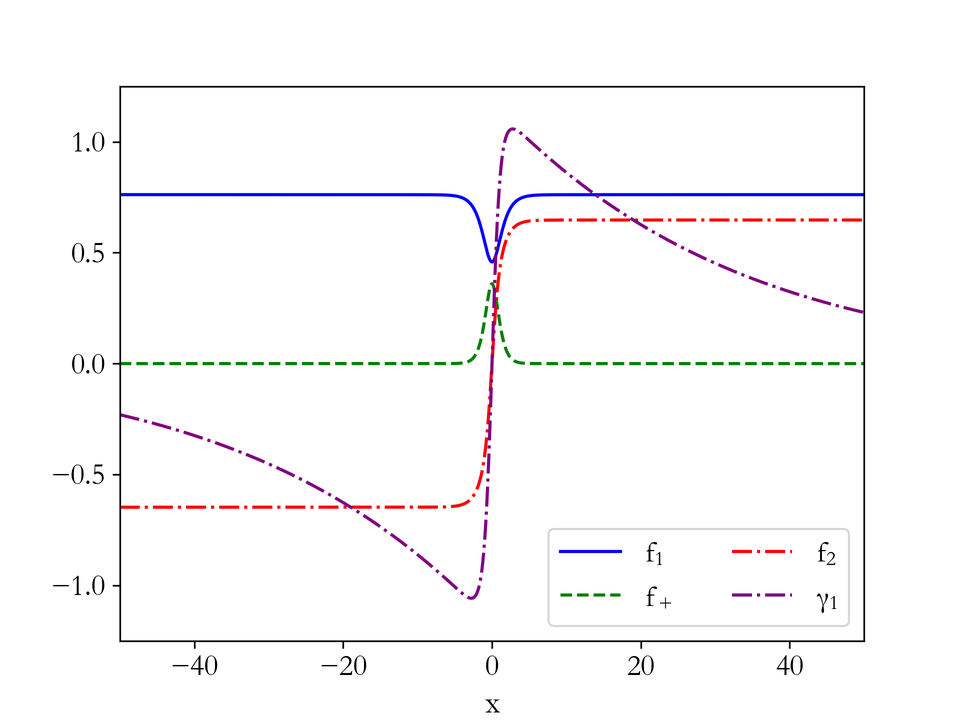}}
    \caption{Example minimum-energy kink solutions for the current-carrying ansatz of (\ref{eq:CC_DW}) for varying values of $\kappa$ in parameter set D. Numerical settings of $n_x = 20000,\, \Delta x = 0.01,\, \delta = 10^{-7}$. Note the standard reduction in the amplitude of the condensate for more negative values of $\kappa$, and the convergence rate of $\gamma_1 \to 0$.}
    \label{fig:CC_mag_sols}
\end{figure*}

In Fig.~\ref{fig:CC_mag_sols} we present kink solutions for this ansatz for varying values of negative $\kappa$.
As is usual for superconducting defects, the condensate amplitude decreases with more negative $\kappa$, we also see the direct influence of $\kappa$ on the convergence of $\gamma_1$.

It is important to mention that current-carrying defect solutions may sometimes be unstable. We therefore analyse when the solutions are predicted to be unstable to simple longitudinal (pinching) and transverse perturbations. These will be apparent in simulations of a perturbed wall, which provides a more robust test of stability.

A semi-analytic analysis based on those detailed in refs.~\cite{LEMPERIERE2003511, CARTER1993151, Battye_Cotterill_2022} for current carrying strings, allows us to predict where such instabilities will occur for a given parameter set. Using the $g_i$ representation of (\ref{eq:CC_DW}) we may write the total energy of the kink solution as
\begin{equation}
    E_{\rm kink} = \int \mathcal{E} dx  = \int (\mathcal{E}_{13} + \mathcal{E}_{24})dx\,,
\label{eq:CC_tot_energy}
\end{equation}
where $\mathcal{E}_{13}$ is the part of the energy density containing the contributions of the fields associated with the current, $g_1$ and $g_3$, while $\mathcal{E}_{24}$ is the remaining energy density, consequently containing terms which only involve the unassociated fields $g_2$ and $g_4$,
\begin{align}
    \mathcal{E}_{13} & = \half (\partial_xg_1)^2 + \half (\partial_xg_3)^2 + \half\left(\omega^2 + k^2\right)\left(g_1^2 + g_3^2\right) \nonumber \\
    & - \half \mu_1^2g_1^2 - \half \mu_2^2g_3^2  \nonumber \\
    & + \quart \lambda_1 g_1^2\left(g_1^2 + 2g_2^2\right) + \quart \lambda_2 g_3^2\left(g_3^2 + 2g_4^2\right) \nonumber \\
    & + \quart \lambda_3 \left(g_1^2g_3^2 + g_1^2g_4^2 + g_2^2g_3^2\right) \nonumber \\
    & + \quart \left(\lambda_4 - |\lambda_5|\right)\left(g_1^2g_3^2 + 2g_1g_2g_3g_4\right)\,,
\end{align}
\begin{align}
    \mathcal{E}_{24} & = \half (\partial_xg_2)^2 + \half (\partial_xg_4)^2 - \half \mu_1^2g_2^2 - \half \mu_2^2g_4^2  \nonumber \\
    & + \quart \lambda_1 g_2^4 + \quart \lambda_2 g_4^4 + \quart \left( \lambda_3 + \lambda_4 - |\lambda_5|\right)g_2^2g_4^2\,.
\end{align}
Taking the equations of motion for $g_1$ and $g_3$, multiplying by $\half g_1$ and $\half g_3$ respectively, integrating each over the cross section of the solution and simplifying the derivatives using integration by parts shows that
\begin{align}
   \int \Bigg[ & \half (\partial_xg_1)^2 -\half\kappa g_1^2 -\half \mu_1^2g_1^2 \nonumber \\
   &+ \half \lambda_1 g_1^2\left(g_1^2 + g_2^2\right) + \quart\lambda_3g_1^2\left(g_3^2 + g_4^2\right) \nonumber \\ 
                &+ \quart\left(\lambda_4 - |\lambda_5|\right)\left(g_1^2g_3^2 + g_1g_2g_3g_4\right) \Bigg]dx = 0\,,
\end{align}
\begin{align}
   \int \Bigg[ & \half (\partial_xg_3)^2 -\half\kappa g_3^2 -\half \mu_2^2g_3^2 \nonumber \\
   &+ \half \lambda_2 g_3^2\left(g_3^2 + g_4^2\right) + \quart\lambda_3g_3^2\left(g_1^2 + g_2^2\right) \nonumber \\ 
                &+ \quart\left(\lambda_4 - |\lambda_5|\right)\left(g_1^2g_3^2 + g_1g_2g_3g_4\right) \Bigg] dx= 0\,.
\end{align}
These can be substituted into the energy (\ref{eq:CC_tot_energy}) such that
\begin{equation}
    E_{\rm kink} = \tau  + \omega^2\Sigma_2\,,
    \label{eq:tot_kink_energy}
\end{equation}
where $\Sigma_2 = \int (g_1^2 + g_3^2) dx$, and 
\begin{align}
    \tau = \int \Bigg[&\mathcal{E}_{24} - \quart\lambda_1g_1^4 - \quart\lambda_2g_3^4 \nonumber \\
          & - \quart \left( \lambda_3 + \lambda_4 - |\lambda_5|\right)g_1^2g_3^3 \Bigg]dx\,.
\end{align}
The Lagrangian may be rewritten using the same technique to yield
\begin{equation}
    L_{\rm kink} = \int \mathcal{L}dx = -\tau\,.
    \label{eq:vort_lag}
\end{equation}
A calculation and diagonalization of the energy momentum tensor, details of which can be found in refs.~\cite{Battye2009SKV,Battye_Cotterill_2022}, shows the propagation speeds of longitudinal and transverse perturbations to be
\begin{equation}
    c_L^2 = 1 + \frac{2\kappa\Sigma_2^\prime}{\Sigma_2}\,, \quad c_T^2 = 1+\frac{\kappa\Sigma_2}{\tau}\,,
    \label{eq:sound_speeds}
\end{equation}
in the magnetic regime, where $\Sigma_2^\prime = \partial \Sigma_2/\partial\kappa$. Both are required to be greater than zero in order for a solution to be stable, in addition to being less than unity to respect causality.

\begin{figure}
      \centering
      \includegraphics[width=\columnwidth]{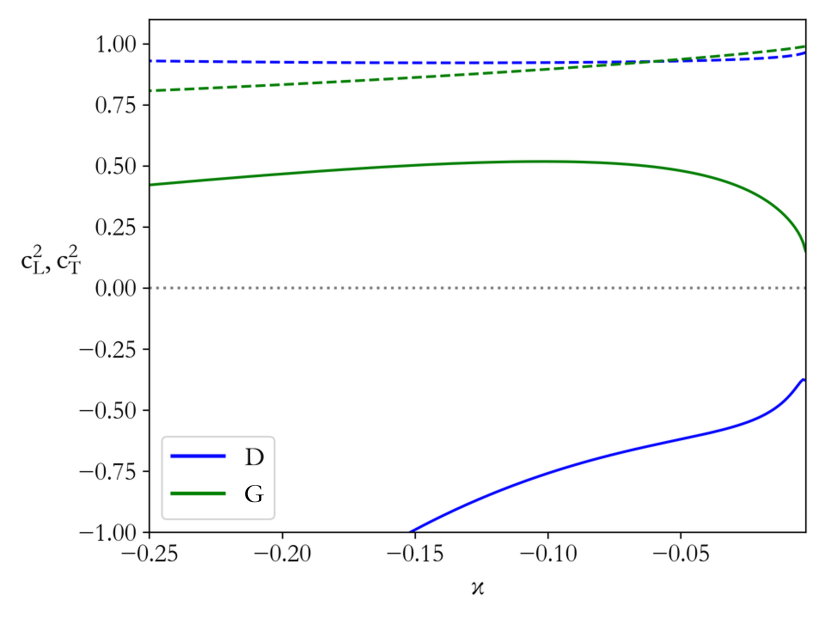}
    \caption{Perturbation propagation speeds, $c_L^2$ and $c_T^2$ for current carrying solutions on the interval $\kappa = [-0.250, -0.001]$, with $\Delta\kappa=0.001$, for two parameter sets using kink solutions with numerical settings of $n_x=8000,\, \Delta x=0.05,\, \delta=10^{-6}$. Solid lines represent $c_L^2$ and dashed $c_T^2$. Parameter set D is predicted to be unstable for the full range of $\kappa$, whereas parameter set G is found to be stable. We find this prediction of stability to be correct.}
    \label{fig:stability_params}
\end{figure}

The variation of these speeds is shown in Fig.~\ref{fig:stability_params} across a range of $\kappa$ for two different parameter sets, with $\Sigma_2$ and $\tau$ having been evaluated for each individual solution on the interval and $\Sigma_2^\prime$ computed using a $2^{nd}$ order finite difference scheme. 
We find that parameter set G is predicted to be stable to such perturbations, whereas parameter set D is not. This analysis demonstrates further that our ansatz is only valid in the magnetic regime for the global theory, as from (\ref{eq:sound_speeds}) it is clear that both speeds should tend to unity as they approach the chiral limit, however from Fig.~\ref{fig:stability_params} we see that this is clearly not the case. This is a direct consequence of the behaviour of $\gamma_1$ in the vacuum and its influence on $\Sigma_2$; as we approach the chiral limit $\gamma_1$ tends to a finite value and as such $\Sigma_2$ diverges at $\kappa=0$. Noteworthy also is that in agreement with ref.~\cite{Battye_Cotterill_2022} we find that $c_L^2 < c_T^2$ for all tested parameter sets.

\begin{figure*}
    \subfloat[$t=0$]{\includegraphics[width=0.15\textwidth]{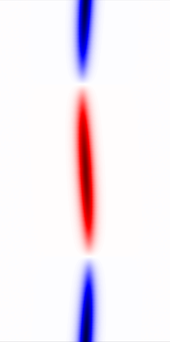}}\hspace{10pt}
    \subfloat[$t=16$]{\includegraphics[width=0.15\textwidth]{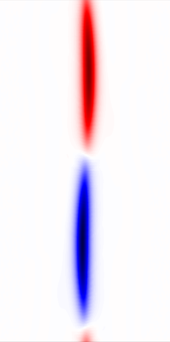}}\hspace{10pt}
    \subfloat[$t=32$]{\includegraphics[width=0.15\textwidth]{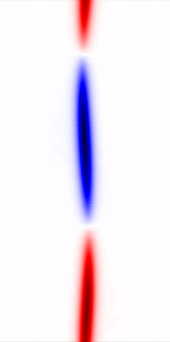}}\hspace{10pt}
    \subfloat[$t=48$]{\includegraphics[width=0.15\textwidth]{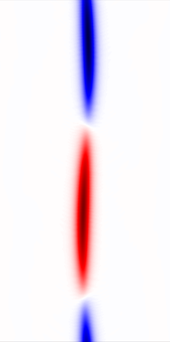}}\hspace{10pt}
    \subfloat[$t=64$]{\includegraphics[width=0.15\textwidth]{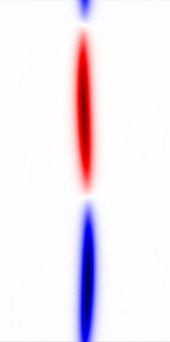}}\hspace{10pt}
    \subfloat{\includegraphics[width=0.0585\textwidth]{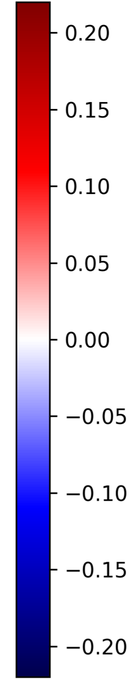}}
\caption{Snapshots of the evolution of the bi-linear component $R^4$ of a magnetic infinite wall solution, within a full dynamical simulation, for parameter set G and $\kappa = -0.025$, the wall has winding number $N=1$ and phase frequency $\omega = 0.116$. The full simulation runs until $t=16000$, with no signs of instability observed.}
\label{fig:inf_wall_evol}
\end{figure*}

To test the stability of our solutions we constructed y-directed domain walls using kink solutions for parameter set G, we set the values of the fields in x by identifying them with those of the relevant kink solution and assigned the phase of the current associated with the coordinates of y and t. These solutions were then evolved in full two-dimensional dynamics in the linear representation. We performed our simulations on rectangular grids of $P_x =4096,\, P_y = 640,\, \Delta x = 0.05,\,\Delta t = 0.01$ (such that the length of the wall is $L=32$) using periodic boundaries in y and homogeneous Neumann boundaries in x. The dimensions of the simulation were chosen to best accommodate the width of the wall and allow for the solution's stability to be tested over a large time period.

In Fig.~\ref{fig:inf_wall_evol} we show snapshots of the early evolution of one of these walls for a value of $\kappa=-0.025$, the wall is initialised with a winding of $N=1$ ($k = 2\pi N / L$), phase frequency of $\omega = 0.116$ and a sinusoidal perturbation of unitary amplitude along the y-direction for a more robust test of stability. We observe the current along the wall length and a coherent oscillation of the entire object induced by the initial perturbation. This current-carrying wall was tested up to $t=16000$ after which the current had performed approximately $300$ full revolutions, and no signs of any instability were observed. We found similar stability for all other walls with values of $\kappa$ where parameter set G is predicted to be stable on the basis of $c^2_L,\, c^2_T > 0$. We also tested current-carrying walls for parameter set D, and found them all to decay due to a growing pinching instability akin to ref.~\cite{Battye:2022mxi}.

 One technical point of interest is that in order for a reliable representation of the dynamics we found a spatial resolution of $\Delta x \leq 0.05$ was required in order for the condensate to remain localised on the wall over longer periods of time. This is likely due to the associated length scales of the solution. 
 
 Of course such simulations do not absolutely confirm that these current-carrying solutions are stable, however given the temporal extent to which we have tested them it is highly suggestive of stability. Given the natural formation of circular current-carrying objects observed in full dynamical simulations \cite{BATTYE2025139311}, we propose that these solutions provide the foundation for constructing Kinky Vortons, a topic currently under investigation which will be presented in future work.

\section{CP-Violating Domain Walls}\label{sec:CP_vio_DW}
In Sec.~\ref{sec:solution} we demonstrated the existence of a new subclass of domain wall solution, of the type shown in Fig.~\ref{fig:eta_1_sol} which can be described by the reduced ansatz of
\begin{equation}
\Phi = \begin{pmatrix}\Phi_1 \\ \Phi_2 \end{pmatrix} = \frac{v_{\rm SM}}{\sqrt{2}}\begin{pmatrix}0 \\ f_1e^{-\half i\eta_1} \\ 0 \\ f_2e^{i\left(\xi-\half \eta_1\right)} \end{pmatrix}\,,
\label{eq:CP_vio_ansatz}
\end{equation}
subject of course to the previous parameter restrictions of $M_{H^\pm} > M_A \quad \text{and} \quad \hat M_A^2 \lesssim \half\left(\hat M_H^2 + f^{-2}_1(0) - 1 \right)$.

\begin{figure*}
    \centering
    \subfloat[]{
        \begin{minipage}[t]{0.33\textwidth}
            \centering
            \includegraphics[width=\columnwidth]{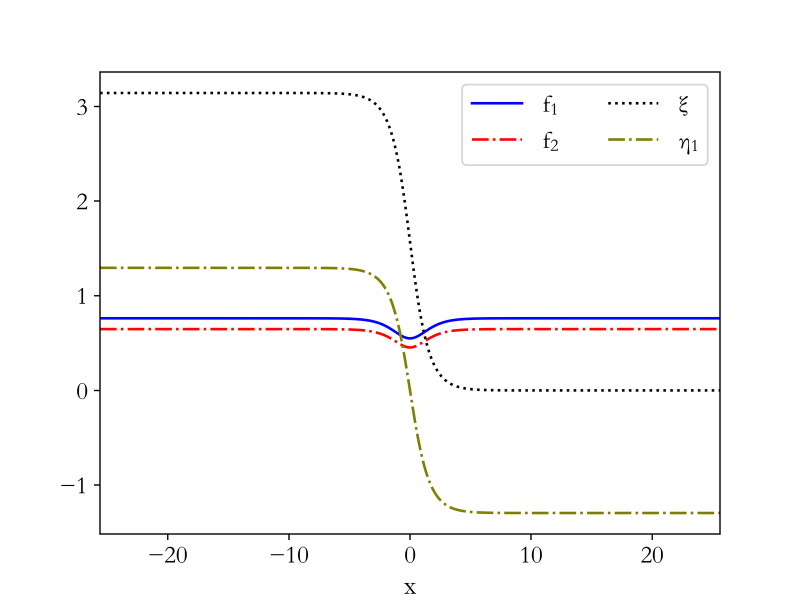}\vspace{4pt} \\
            \includegraphics[width=\columnwidth]{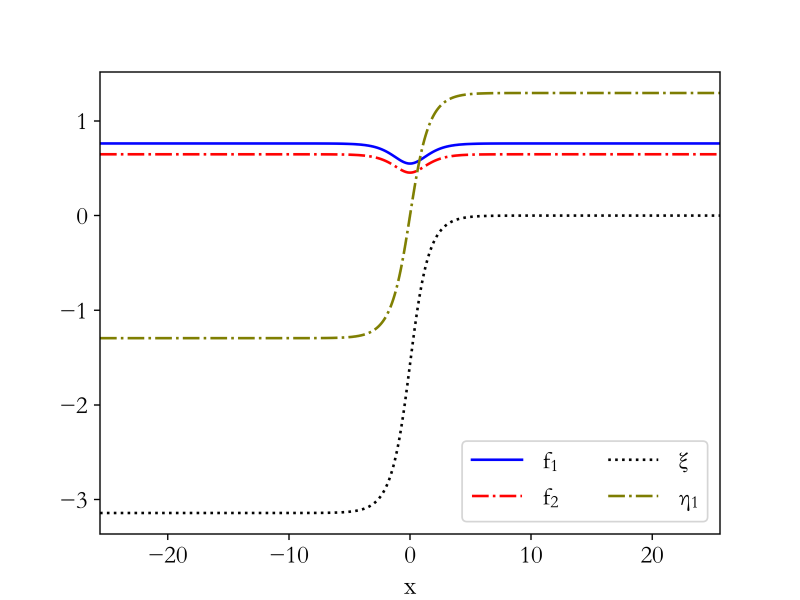}\vspace{4pt}
            \label{fig:CP_vio_gen}
        \end{minipage}
    }
    \subfloat[]{
        \begin{minipage}[t]{0.33\textwidth}
            \centering
            \includegraphics[width=\columnwidth]{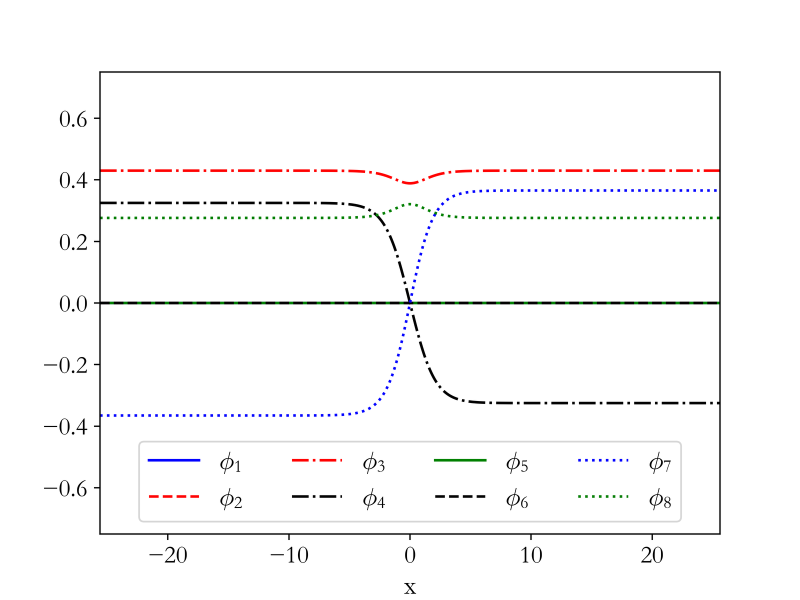}\vspace{4pt}\\
            \includegraphics[width=\columnwidth]{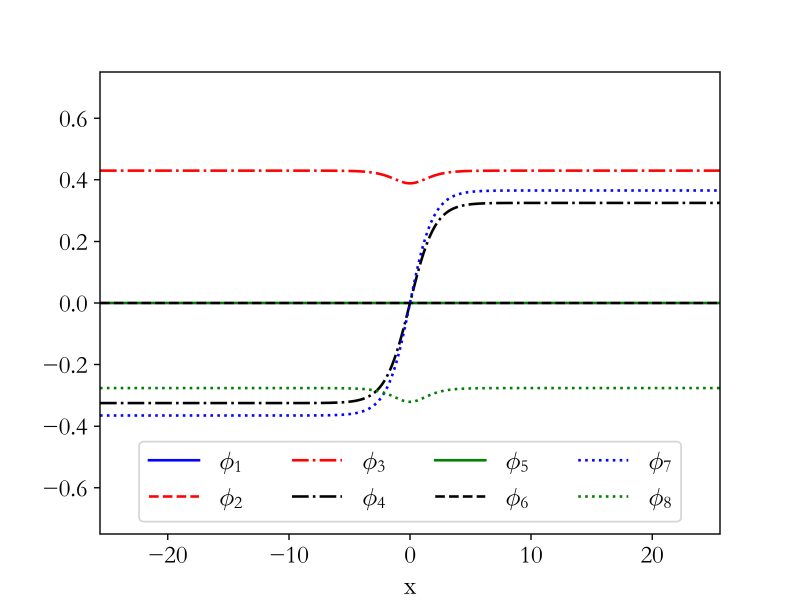}\vspace{4pt}
            \label{fig:CP_vio_phi}
        \end{minipage}
    }
    \subfloat[]{
        \begin{minipage}[t]{0.33\textwidth}
            \centering
            \includegraphics[width=\columnwidth]{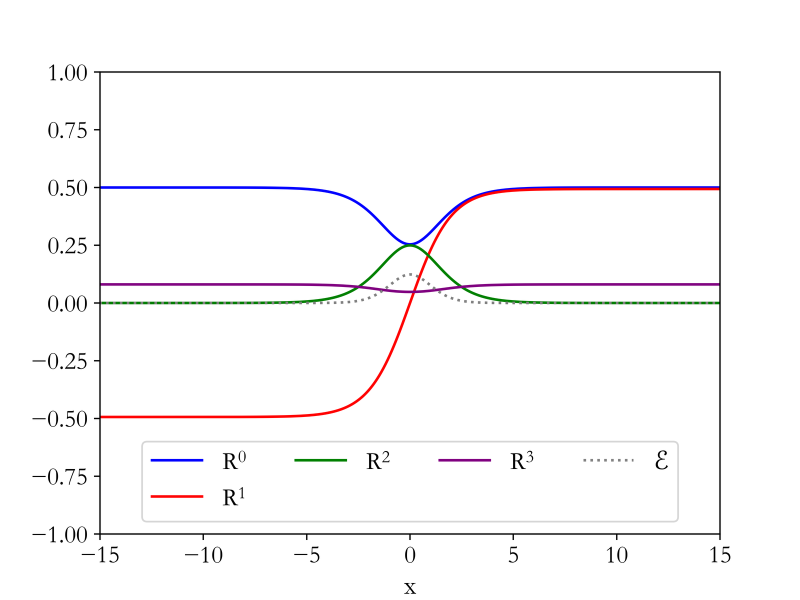}\vspace{4pt}\\
            \includegraphics[width=\columnwidth]{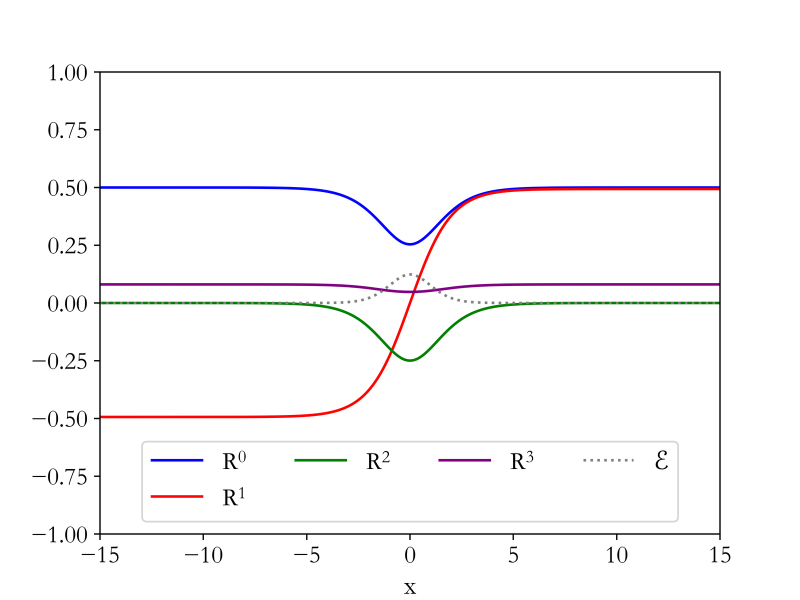}\vspace{4pt}
            \label{fig:CP_vio_R}
        \end{minipage}
    }
    \caption{Minimum-energy kink solutions for parameter set H, depicting fields in the (a) general representation, (b) linear representation (middle) and (c) bi-linear representation (bottom) for two equivalent CP-violating solutions. In the linear representation, $\phi_i$ for $i=1,\,2,\,5,\,6$ are globally zero. Numerical settings of $n_x = 5000, \Delta x = 0.01, \delta = 10^{-7}$.}
    \label{fig:CP_vio_1D} 
\end{figure*}

This reduced ansatz admits two domain wall solutions, related by a CP1 transformation, $\Phi \rightarrow \Phi^*$, shown in Fig.~\ref{fig:CP_vio_1D}. They differ only by the sign of $R^2$ across the wall, while their vacuum $R^\mu$ configurations remain identical. We have used the further parameter set H as it resides well into the mass regime of these solutions.
Motivated by this, and by observations from full dynamical simulations in this mass regime, we propose that a CP1 domain wall can form longitudinally upon the underlying $\mathbb{Z}_2$ wall.

To test this hypothesis, we constructed a two-dimensional grid with the $\mathbb{Z}_2$ wall aligned along the $y$-axis. Opposite sides of the grid were initialized with the two CP-violating solutions of Fig.~\ref{fig:CP_vio_phi}, with a thin interpolating section (of $\tanh$ form) at the centre. This served as the initial condition for solving the two-dimensional equations of motion with homogeneous Neumann boundary conditions in both $x$ and $y$, ensuring relaxation to a minimum-energy configuration.

\begin{figure*}
    \centering
    \subfloat[$R^1$]{
        \begin{minipage}[b]{0.33\textwidth}
            \centering
            \includegraphics[width=0.75\columnwidth]{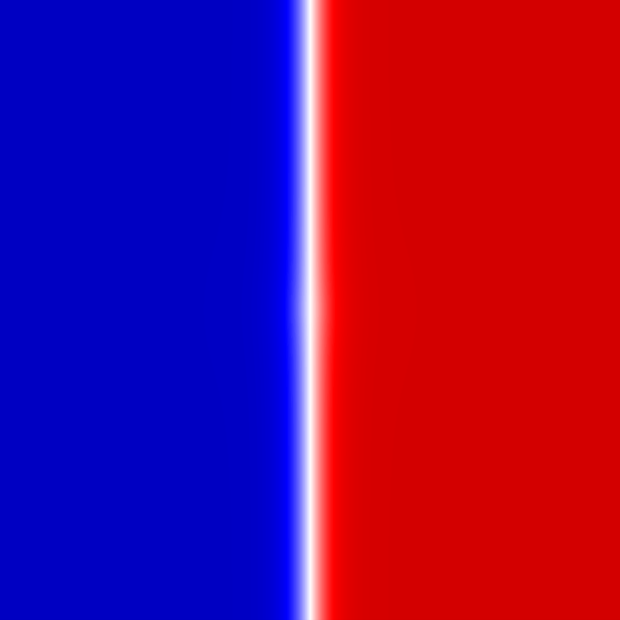}
            \includegraphics[width=0.11\columnwidth]{RICcb.png}
            \label{fig:CP_vio_2d_R1}
        \end{minipage}
    }
    \subfloat[$R^2$]{
        \begin{minipage}[b]{0.33\textwidth}
            \centering
            \includegraphics[width=0.75\columnwidth]{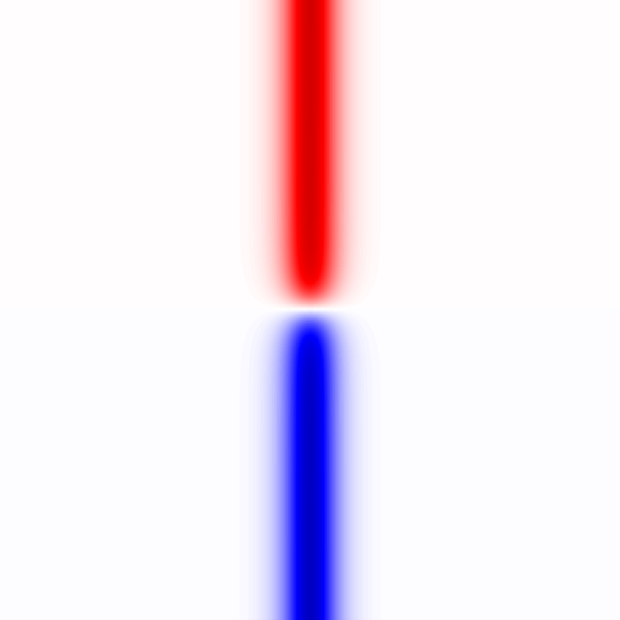}
            \includegraphics[width=0.11\columnwidth]{RICcb.png}
            \label{fig:CP_vio_2d_R2}
        \end{minipage}
    }
    \subfloat[$R^\mu$ (longitudinal)]{
        \begin{minipage}[b]{0.33\textwidth}
            \centering
            \includegraphics[width=\columnwidth]{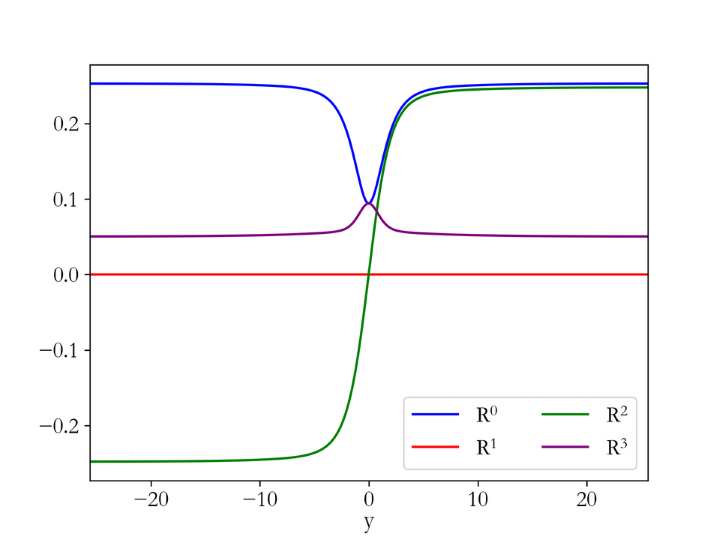}
            \label{fig:CP_vio_2d_wall_fields}
        \end{minipage}
    }
    \caption{Stable two-dimensional domain wall solution for parameter set H, where initial conditions were constructed such that a CP1 domain wall existed upon the $\mathbb{Z}_2$ domain wall. We see this two-dimensional configuration to be stable, with maximal CP-violation occurring longitudinally upon the domain wall.}
    \label{fig:2d_cp_vio}
\end{figure*}

We present in Fig.~\ref{fig:2d_cp_vio} a fully converged (to a tolerance of $10^{-7}$) two-dimensional domain wall solution in the bi-linear components $R^1$ and $R^2$, along with the full bi-linear component profiles along the domain wall. The converged solution clearly shows a superposition, where a CP1 domain wall exists upon a $\mathbb{Z}_2$ domain wall. The longitudinal profiles of $R^\mu$ directly correspond to the forms of the CP1 kink solution found in ref.~\cite{Battye2021SDW}. Explicitly, this solution realises local maximal CP violation confined to the $\mathbb{Z}_2$ domain wall.

\begin{figure*}
    \subfloat[$t=0$]{\includegraphics[width=0.15\textwidth]{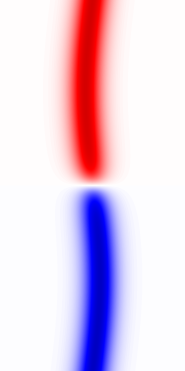}}\hspace{10pt}
    \subfloat[$t=32$]{\includegraphics[width=0.15\textwidth]{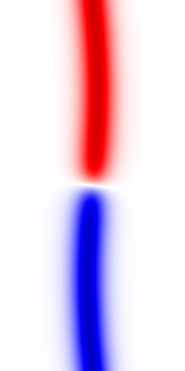}}\hspace{10pt}
    \subfloat[$t=64$]{\includegraphics[width=0.15\textwidth]{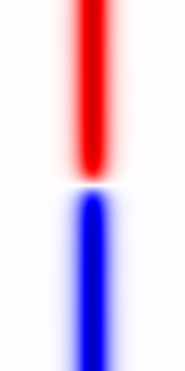}}\hspace{10pt}
    \subfloat[$t=96$]{\includegraphics[width=0.15\textwidth]{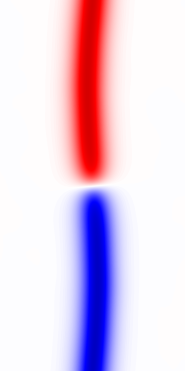}}\hspace{10pt}
    \subfloat[$t=128$]{\includegraphics[width=0.15\textwidth]{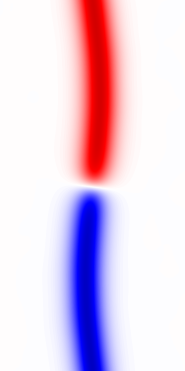}}\hspace{10pt}
    \subfloat{\includegraphics[width=0.0558\textwidth]{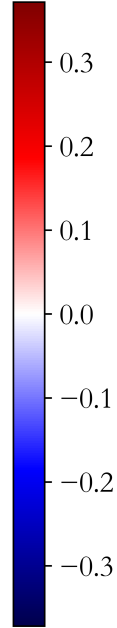}}
  \caption{Snapshots of the evolution, within a full dynamical simulation, of the bi-linear component $R^2$ of a composite domain wall structure where a CP1 wall exists longitudinally upon a $\mathbb{Z}_2$ wall, initialised in a sinusoidally perturbed state. No signs of instability were observed.}
  \label{fig:2d_cp_vio_sim}
\end{figure*}

To test the stability of this composite domain wall structure we took the solution which we have presented in Fig.~\ref{fig:2d_cp_vio} and evolved it within a full dynamical simulation from a sinusoidally perturbed initial state, using numerical settings of $P_x =1024,\, \Delta x = 0.05,\,\Delta t = 0.01$ and Neumann boundaries in both spatial dimensions. We show in Fig~\ref{fig:2d_cp_vio_sim} the evolution of this simulation, where we see an oscillation of the entire object due to the initial perturbation, with the CP1 wall persisting upon the $\mathbb{Z}_2$ wall under the full dynamics and exhibiting no signs of instability. While this simulation is only a short test of stability, it demonstrates that the composite structure does indeed persist under full dynamical evolution, suggesting stability of the solution.

The existence of this stable field configuration completes the explanation of the features found to exist within full dynamical $(2+1)$-dimensional simulations of this model. That is to say, that the previously observed ``winding" in $R^2$ is in fact the formation of CP1 domain walls upon those which naturally occur in the $\mathbb{Z}_2$-symmetric model.

Although speculative, the existence of such solutions has far-reaching cosmological implications. Since domain walls must ultimately decay to avoid over-closing the universe, the presence of local CP violation along them would naturally create an out-of-equilibrium environment, potentially conducive to baryogenesis. Unlike the ``winding" of a condensate on a two-dimensional wall (a proxy for the winding of a three dimensional string), this superposition of domain walls can be straightforwardly extended to three dimensions, strengthening its cosmological relevance.

\section{Conclusions}\label{sec:conclusion}
The underlying objective of this work was to explain the phenomena observed in full dynamical, two-dimensional simulations of the $\mathbb{Z}_2$-symmetric 2HDM. To this end, we have presented a comprehensive study of domain wall solutions in this model, identifying two broad classes of solutions: (i) those connecting vacua related by an electroweak rotation, and (ii) those corresponding to distinct energy-minimising configurations. Together, these two classes provide a comprehensive explanation of the simulation dynamics.

We have categorised the model’s parameter space into four distinct regions, in which different subclasses of solution are the energy-minimising field configuration. We summarise these solutions below in terms of the scaled mass parameters $\hat M_H$, $\hat M_{H^\pm}$, $\hat M_A$, and $\tan\beta$,
\begin{itemize}
\item \textbf{Standard solutions} (cf.~Sec.\ref{sec:standard_sol}) occur when:
\begin{align*}
\hat M_{\min}^2 \gtrsim \half\left(\hat M_H^2 +f^{-2}_1(0) - 1 \right)\,, 
\end{align*}
where $\hat M_{\min} = \min(\hat M_{H^\pm},\, \hat M_A)$. These solutions are fully described by the field components $f_1$ and $f_2$.

\item \textbf{Superconducting solutions} (cf.~Sec.\ref{sec:gamma1}) arise when:
\begin{align*}
    \hat M_{H^\pm} < \hat M_A \quad \text{and} \quad \hat M_{H^\pm}^2 \lesssim \half\left(\hat M_H^2 +f^{-2}_1(0) - 1 \right).
\end{align*}
These are characterised by the field components $f_1$, $f_+$, $f_2$, and the $SU(2)$ parameter $\gamma_1$.

\item \textbf{CP-violating solutions} (cf.~Sec.\ref{sec:eta1}) appear when:
\begin{align*}
    \hat M_A < \hat M_{H^\pm} \quad \text{and} \quad \hat M_A^2 \lesssim \half\left(\hat M_H^2 +f^{-2}_1(0) - 1 \right).
\end{align*}
These are described by the field components $f_1$, $f_2$, $\xi$, and the $SU(2)$ parameter $\eta_1$.

\item \textbf{Simultaneously superconducting \& CP-violating solutions} occur when:
\begin{align*}
    \hat M_A = \hat M_{H^\pm} \quad \text{and} \quad \hat M_A^2 \lesssim \half\left(\hat M_H^2 +f^{-2}_1(0) - 1 \right).
\end{align*}
These involve the full set of field components: $f_1$, $f_+$, $f_2$, $\xi$, and the $SU(2)$ parameters $\gamma_1$, $\eta_1$.
\end{itemize}
Together with the effects of relatively EW transformed vacua, the above solutions provide a complete explanation of the structures observed in $(2+1)$-dimensional simulations of the model from random initial conditions. The excellent agreement between this theoretical classification and full dynamical simulations gives us confidence that the relevant minimum-energy configurations have been identified. 
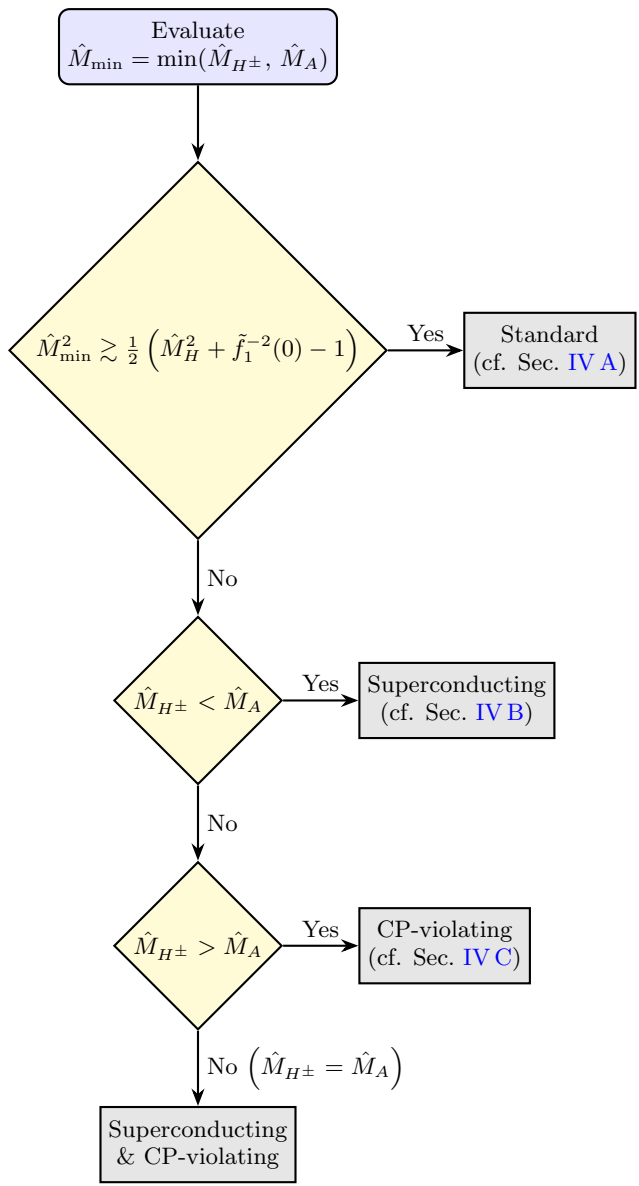
\begin{figure}
    \tikzstyle{decision} = [diamond, draw=black, thick, fill=yellow!20,
                            minimum width=1cm, minimum height=1cm, inner sep=1pt, text centered]
    \tikzstyle{process}  = [rectangle, draw=black, thick, fill=gray!20,
                            minimum width=1cm, minimum height=1cm, text centered]
    \tikzstyle{startstop} = [rectangle, rounded corners, draw=black, thick, fill=blue!10,
                             minimum width=1cm, minimum height=1cm, text centered]
    \tikzstyle{arrow} = [thick, ->, >=Stealth]
    
    \begin{center}
    \begin{tikzpicture}[node distance=1cm and 1cm]
    
    \node (start) [startstop] {\begin{varwidth}{5cm}\centering Evaluate\\ $\hat M_{\min} = \min(\hat M_{H^\pm},\, \hat M_A)$\end{varwidth}};
    \node (cond1) [decision, below=1cm of start] {$\hat M_{\min}^2 \gtrsim \half\left(\hat M_H^2 +\tilde f^{-2}_1(0) - 1 \right)$};
    \node (standard) [process, right=1cm of cond1] {\begin{varwidth}{5cm}\centering Standard \\ (cf. Sec.~\ref{sec:standard_sol})\end{varwidth}};
    
    \node (cond2) [decision, below=of cond1] {$\hat M_{H^\pm} < \hat M_A$};
    \node (sc) [process, right=1cm of cond2] {\begin{varwidth}{5cm}\centering Superconducting \\ (cf. Sec.~\ref{sec:gamma1})\end{varwidth}};
    
    \node (cond3) [decision, below=of cond2] {$\hat M_{H^\pm} > \hat M_A$};
    \node (cpv) [process, right=1cm of cond3] {\begin{varwidth}{5cm}\centering CP-violating \\ (cf. Sec.~\ref{sec:eta1})\end{varwidth}};
    
    \node (both) [process, below=1cm of cond3, fill=gray!20] {\begin{varwidth}{5cm}\centering Superconducting\\\& CP-violating\end{varwidth}};

    \draw [arrow] (start) -- (cond1);
    \draw [arrow] (cond1.east) -- (standard.west) node[midway, above] {Yes};
    \draw [arrow] (cond1) -- (cond2) node[midway, right] {No};
    
    \draw [arrow] (cond2.east) -- (sc.west) node[midway, above] {Yes};
    \draw [arrow] (cond2) -- (cond3) node[midway, right] {No};
    
    \draw [arrow] (cond3.east) -- (cpv.west) node[midway, above] {Yes};
    \draw [arrow] (cond3) -- (both) node[midway, right] {No $\left(\hat M_{H^\pm} = \hat M_A\right)$};
    
    \end{tikzpicture}
    \end{center}
    \caption{Flowchart which, given a parameter set, will yield the approximate domain wall solution type in the $\mathbb{Z}_2$-symmetric 2HDM. The quantity $f_1(0)$ may be evaluated directly using the reduced ansatz of the standard solution for the best prediction, using (\ref{eq:sol_region_eval}) for a robust approximate prediction, or (\ref{eq:sol_region_approx}) close to $\hat M_H^2 = 1$. This classification remains an approximation due to the neglecting of gradient energies.}
    \label{fig:flowchart}
\end{figure}
A concise delineation of the parameter space as we have outlined can be found in Fig.~\ref{fig:flowchart}, a flowchart mapping any given parameter set combinations to the approximate solution subclass.

As in our previous work on superconducting strings in the 2HDM~\cite{Battye:2024dvw}, we have demonstrated that the superconducting subclass of domain wall solutions can support persistent currents, resulting in apparently stable current-carrying walls in two dimensions. These walls could serve as the basis for constructing closed-loop configurations, so-called “Kinky Vortons”, in the 2HDM, which would be the $(2+1)$-dimensional analogues of Vortons in the $U(1)$-symmetric variant of the theory. If such objects can be successfully constructed, they would provide a valuable lower-dimensional test of vorton properties within this model, offering insight into their stability in a far more computationally tractable setting.

Finally, we have discovered a new type of domain wall solution in the 2HDM: a CP-violating wall configuration in which CP symmetry is maximally locally broken along the $\mathbb{Z}_2$-symmetric domain wall. Unlike the current-carrying walls (which primarily serve as proxies for current-carrying strings in three dimensions), this CP-violating wall solution should have an analogue in higher dimensions as well. In realistic scenarios, domain walls must ultimately decay to avoid cosmological problems; notably, the presence of CP violation along such unstable walls means they naturally fulfil two of Sakharov’s three conditions (CP violation and departure from equilibrium) for generating a matter–antimatter asymmetry~\cite{Sakharov:1967dj}. Whether these walls can actually produce a net baryon asymmetry depends on the details of how they interact with Standard Model fermions, and further work will be necessary to determine if these CP-violating walls can efficiently fuel baryogenesis.

In summary, this study has elucidated the rich spectrum of domain wall solutions in the 2HDM and linked our theoretical classification of these solutions with their observed dynamics in simulations. We have provided a complete categorisation of the minimum-energy field configurations and identified the conditions under which each subclass occurs, thereby offering a predictive roadmap for domain wall phenomena in this model. These results have significant cosmological implications: by understanding which domain wall solutions occur - and whether they are stable or metastable - we gain insight into the possible relics of the electroweak phase transition and their roles in the early Universe. For example, the existence of current-carrying walls and the prospect of Kinky Vortons offer a computationally tractable way to explore vorton dynamics in the 2HDM. Meanwhile, the CP-violating walls introduce a novel mechanism that could be relevant to baryogenesis. We conclude that the 2HDM, a well-motivated extension of the Standard Model, supports a far more varied set of topological phenomena than previously recognised. Future work will focus on exploring the viability of Kinky Vortons, examining how these domain wall solutions interact with particles, and investigating their three-dimensional extensions in realistic cosmological settings. Our findings broaden the theoretical groundwork for such studies and open several new avenues at the intersection of particle physics and cosmology.

\bibliography{References}

\appendix
\section{Numerical Techniques}\label{sec:numericals}
\subsection{One-Dimensional Kink Solutions}\label{sec:numericals_kinks}
\par One-dimensional, static, kink solutions, appropriate for the study of domain walls, can be obtained by using variational techniques to solve the set of $2^{nd}$ order differential equations of motion,
\begin{equation}
    \frac{d}{dx} \left(\frac{\partial \mathcal{L}}{\partial \left(dq/dx \right)} \right) - \frac{\partial \mathcal{L}}{\partial q} = 0\,,
\end{equation}
where $q$ is the set of field functions of a given representation, e.g. $q=\{\phi_1, ..., \phi_8 \}$. The general form of the Lagrangian density for the 2HDM is given by, in one spatial dimension,
\begin{equation}
\mathcal{L} = -\frac{d\Phi_1^\dagger}{dx}\frac{d\Phi_1}{dx} - \frac{d\Phi_2^\dagger}{dx}\frac{d\Phi_2}{dx} - V(\Phi_1, \Phi_2)\,.
\end{equation}
These solutions correspond to static solutions to the equations of motion, subject to the boundary conditions imposed. The solutions are solved for on one-dimensional arrays of size $n_x$, grid spacing of $\Delta x$ subject to a tolerance of $\delta$. 

\subsubsection{Approximate Treatment}
Here we implement a method similar to the variational technique of successive over relaxation (SOR) where functions of each field variable, $q^i$ are introduced,
\begin{equation}
    F_1(q^i) = \frac{d}{dx} \left(\frac{\partial \mathcal{L}}{\partial \left(dq^i/dx \right)} \right) - \frac{\partial \mathcal{L}}{\partial q^i}, \qquad F_2(q^i) = \frac{\partial F_1(q^i)}{\partial q^i}\,.
\end{equation}
We start with an initial trial field configuration (often a linear interpolation between the two boundary values). Then, at each iteration $n$, we simultaneously update all field variables $q^i$ according to
\begin{equation}
    q^i_{n+1} = q^i_n - w \frac{F_1(q^i_n)}{F_2(q^i_n)}\,,
\end{equation}
until all changes are below a tolerance $\delta$. When implementing this technique we use $w=1.5$. We applied either Neumann or Dirichlet boundary conditions, as appropriate for the solution type. Spatial derivatives were evaluated with a fourth-order finite difference stencil.

\subsubsection{Full Treatment}
Where the equations of motion were highly coupled, for example in the cases detailed in Sec.~\ref{sec:solution}, a more robust treatment for solving the one-dimensional equations of motion was required for stable convergence. This was done by implementing a modified Newton's method approach with iterative updates,
\begin{equation}
    {\bf q}_{n+1} = {\bf q}_n - w {\bf J}_n^{-1}{\bf F}_n\,,
\end{equation}
where,
\begin{eqnarray}
    {\bf q}=\begin{pmatrix}
        q^0\\.\\.\\.\\q^i
    \end{pmatrix}\,, \quad
    {\bf F} = \begin{pmatrix}
        F_1(q^0)\\.\\.\\.\\F_1(q^i)
    \end{pmatrix}\,, \nonumber\\\nonumber\\
    {\bf J} =\begin{pmatrix}
        \frac{\partial F_1(q^0)}{\partial q^0} & . &. & .& \frac{\partial F_1(q^0)}{\partial q^i}\\ .&.&&&.\\.&&.&&.\\.&&&.&.\\\frac{\partial F_1(q^i)}{\partial q^0}&.&.&.&\frac{\partial F_1(q^i)}{\partial q^i}
    \end{pmatrix}\,.
\end{eqnarray}
Generally where this method was implemented the equations were very stiff and thus required the use of a lower relaxation coefficient of $w<1$ in order for stable convergence. In addition, derivatives were approximated to $6^{th}$ order for improved accuracy.

\subsection{Full Dynamical Simulations}\label{sec:numericals_dyn}
$(2+1)$-dimensional simulations were performed on regular square grids of $P^2$ data points with grid spacing $\Delta x$ and time-step $\Delta t$ or in the case of our infinite domain wall solutions rectangular grids of $P_x \times P_y$ grid points. The equations of motion were discretised using central finite difference methods with temporal derivatives calculated to $2^{nd}$ order and spatial derivatives to $4^{th}$ order. Neumann or periodic boundary conditions were used at the boundaries of the simulation grid. These full dynamical simulations are performed in the linear representation of (\ref{eq:lin_rep}).

\subsection{Dimensionless Rescaling}\label{sec:dim_rescaling}
Within numerical studies it is convenient to rescale the potential for dimensionless length and energy. Given that the values of $M_h \text{ and } v_{\rm SM}$ are fixed by experiment, these quantities are used for the rescaling as follows,
\begin{eqnarray}
    \Phi \rightarrow v_{\rm SM} \Phi\,, \quad x \rightarrow M_h^{-1}x\,,
\end{eqnarray}
which results in
\begin{eqnarray}
\lambda_{i} \rightarrow \frac{v^2_{SM}}{M^2_h}\lambda_{i}\,, \qquad \mu^2_{j} \rightarrow \frac{1}{M^2_h}\mu^2_{j}\,,
\end{eqnarray}
such that the model parameters only depend on the mass ratios. This rescaling is such that lengths are expressed in units of $M_h^{-1}$ and energy densities in units of $v_{\rm SM}^2M^2_h$. These two parameters can then subsequently be set to unity to achieve a simple numerical scale.

\section{Vanishing Fields}\label{sec:vanishing_fields}
In the regions of parameter space corresponding to each solution subclass, any field that is not part of that subclass’s ansatz indeed vanishes identically in the minimum-energy solution. In ref.~\cite{Sassi:2023cqp}, it was proposed that under variation of all eight general field components in their parameter set, which under our classification would have a superconducting solution, there remained a small but non-zero CP-violating phase $\xi$ at the wall, in addition to minimal non-zero profiles in the other fields of their chosen representation: we suspect these were numerical artifacts.

\begin{figure*}
  \centering
    \subfloat[Parameter set B]{\includegraphics[width=0.5\linewidth]{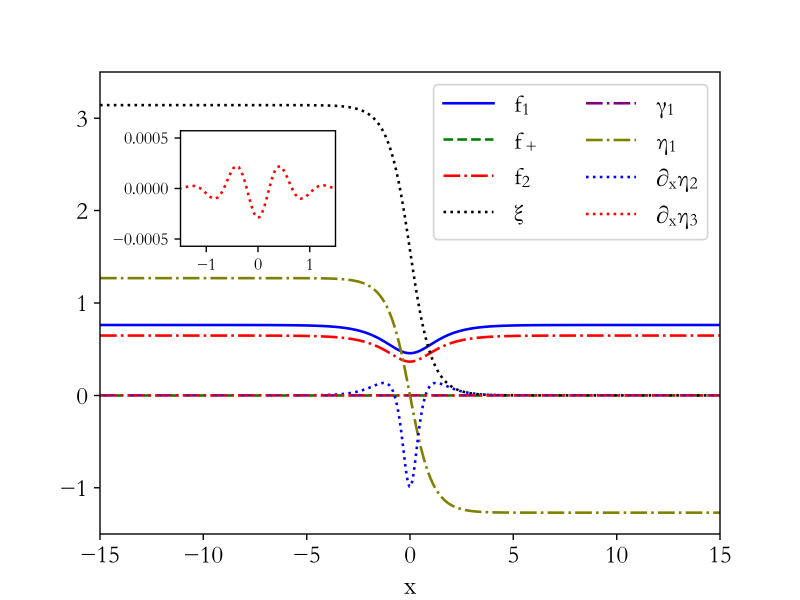}\label{fig:gen_sol_fields_623}}
    \subfloat[Parameter set C]{\includegraphics[width=0.5\linewidth]{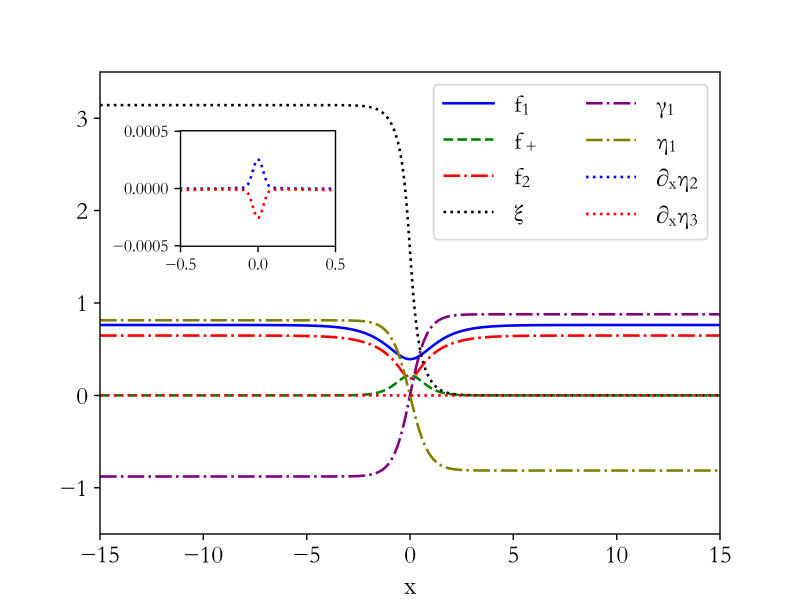}\label{fig:gen_sol_fields_633}}\\
    \subfloat[Parameter set D]{\includegraphics[width=0.5\linewidth]{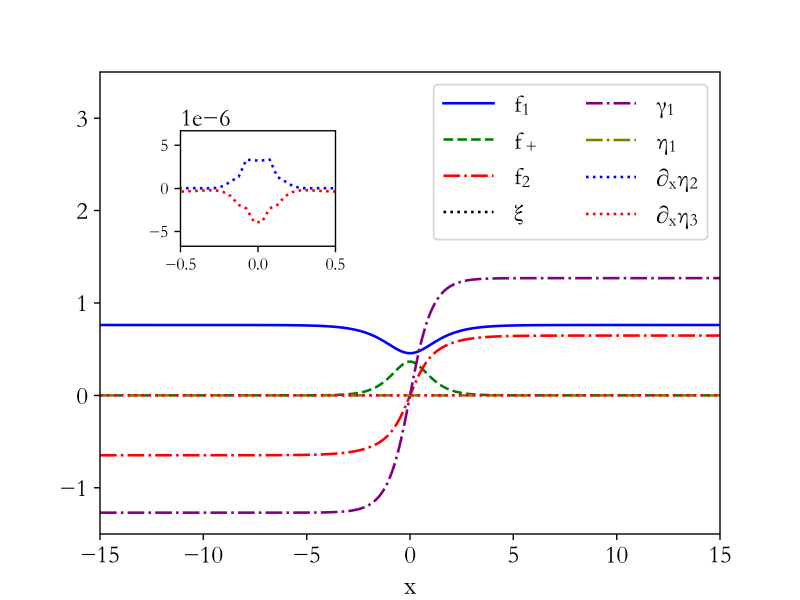}\label{fig:gen_sol_fields_632}}
    \subfloat[Parameter set D]{\includegraphics[width=0.5\linewidth]{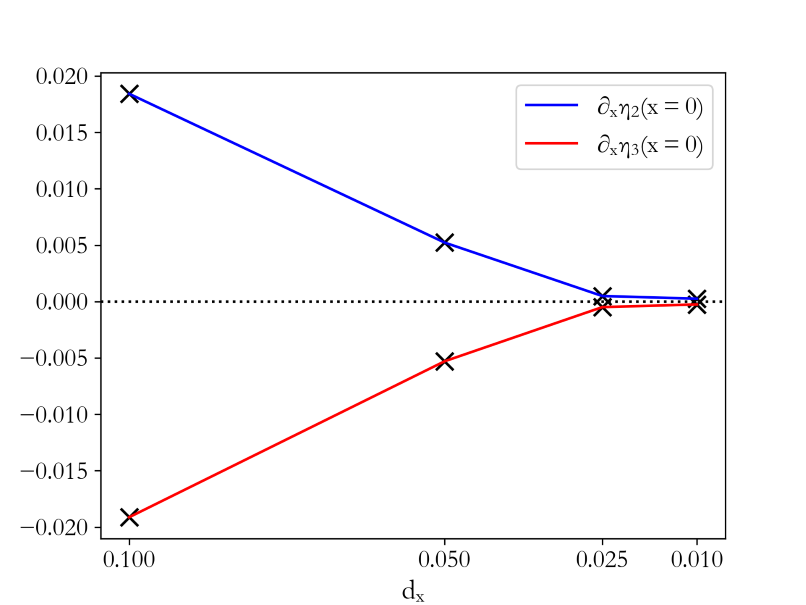}\label{fig:gen_sol_tol_reduction}}
    \caption{(a, b, c) Minimum-energy domain wall solutions in the $\mathbb{Z}_2$-symmetric 2HDM, for three different parameter sets. (d) Convergence towards zero of the general solution quantities $\partial_x\eta_2$ and $\partial_x\eta_3$ for minimum-energy solutions where $M_{H^\pm} \leq M_A$. Their value at the centre of the kink solution is shown as a function of the spatial resolution $d_x$ of the numerical solution. Note that $d_x$ decreases to the right.}
    \label{fig:gen_sols}
\end{figure*}

Further to this point, we show in Fig.~\ref{fig:gen_sols} minimum-energy solutions for three different parameter sets with contrasting solutions. We see that all of the solutions are correctly identified by our classification. Included in the inset plots are the quantities $\partial_x\eta_2$ and $\partial_x\eta_3$, which we used to reduce the field configuration so that it is described by only six functions. As can be seen, in all solutions $\partial_x\eta_3$ is effectively zero in all cases, as is $\partial_x\eta_2$ if the solution has a condensate. We show in Fig.~\ref{fig:gen_sol_tol_reduction} that this minimal profile in both of these quantities is a numerical artifact of the resolution of a solution.

For CP-violating solutions, $\partial_x\eta_2$ does not fully vanish even at high resolution. However, this seems to be entirely a numerical effect: if we consider only the non-vanishing fields ($f_1, f_2, \xi, \eta_1$) in a CP-violating solution, one can show that the equations of motion are satisfied regardless of $\partial_x\eta_2$’s behaviour. We have verified this by testing many parameter sets beyond those shown here. This confirms that our subclasses of solutions successfully describe the full parameter space of the model, meaning that in each solution regime, the minimum-energy configuration involves only the subset of fields we identified for that regime (all other fields are exactly zero). Furthermore, while we do see a tiny non-zero CP-violating phase within the superconducting solution regime, this was found to follow the same behaviour as $\partial_x \eta_2$ and $\partial_x\eta_3$ in that it reduces for smaller grid spacings and smaller solution tolerance.

\section{Further Minimum-Energy Solution Examples}\label{sec:gen_sols_set}
\begin{figure*}
\centering
    \subfloat[$M_H = 200$, $M_A = 200$, $M_{H^\pm} = 200$, $\tan\beta=0.85$]{\includegraphics[width=0.32\textwidth]{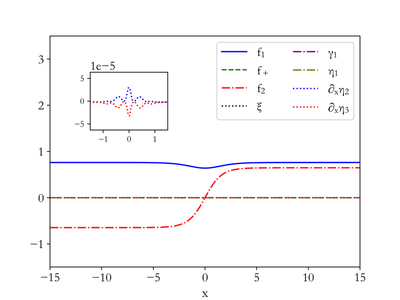}\label{fig:gen_sol_222_0.85}}
    \subfloat[$M_H = 800$, $M_A = 500$, $M_{H^\pm} = 400$, $\tan\beta=0.85$]{\includegraphics[width=0.32\textwidth]{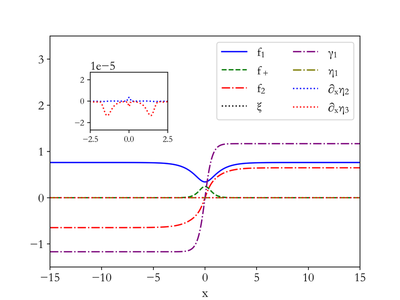}\label{fig:gen_sol_854_0.85}}
    \subfloat[$M_H = 700$, $M_A = 700$, $M_{H^\pm} = 700$, $\tan\beta=0.85$]{\includegraphics[width=0.32\textwidth]{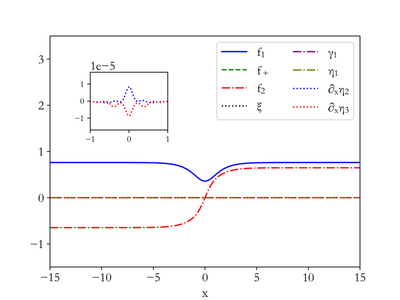}\label{fig:gen_sol_777_0.85}}\\
    \subfloat[$M_H = 650$, $M_A = 300$, $M_{H^\pm} = 500$, $\tan\beta=0.85$]{\includegraphics[width=0.32\textwidth]{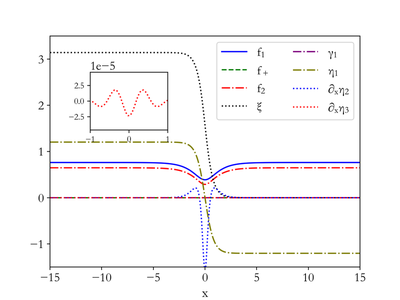}\label{fig:gen_sol_6535_0.85}}
    \subfloat[$M_H = 600$, $M_A = 600$, $M_{H^\pm} = 300$, $\tan\beta=0.85$]{\includegraphics[width=0.32\textwidth]{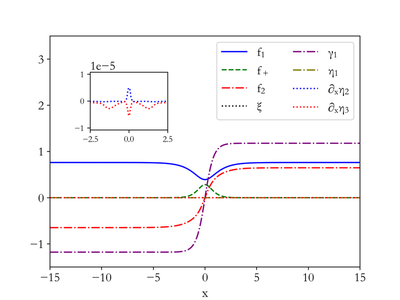}\label{fig:gen_sol_663_0.85}}
    \subfloat[$M_H = 500$, $M_A = 600$, $M_{H^\pm} = 700$, $\tan\beta=0.85$]{\includegraphics[width=0.32\textwidth]{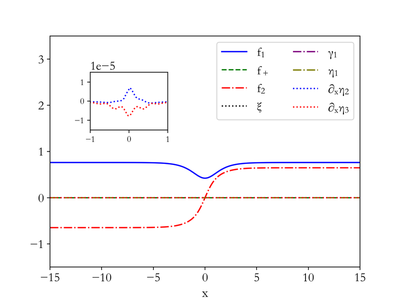}\label{fig:gen_sol_567_0.85}}\\
    \subfloat[$M_H = 400$, $M_A = 700$, $M_{H^\pm} = 600$, $\tan\beta=0.85$]{\includegraphics[width=0.32\textwidth]{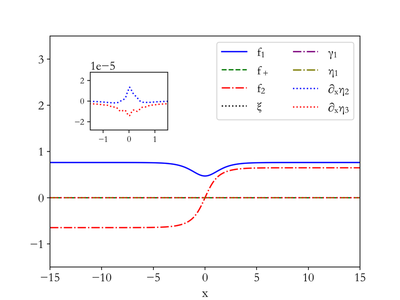}\label{fig:gen_sol_476_0.85}}
    \subfloat[$M_H = 500$, $M_A = 150$, $M_{H^\pm} = 150$, $\tan\beta=0.50$]{\includegraphics[width=0.32\textwidth]{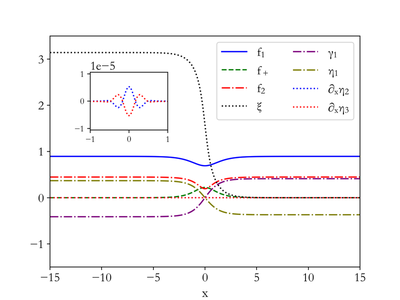}\label{fig:gen_sol_51515_0.50}}
    \subfloat[$M_H = 500$, $M_A = 500$, $M_{H^\pm} = 200$, $\tan\beta=0.50$]{\includegraphics[width=0.32\textwidth]{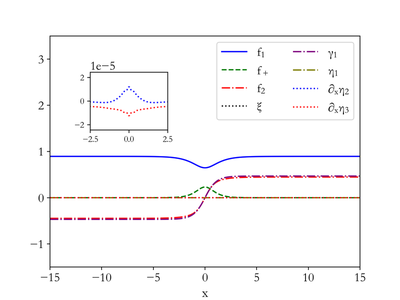}\label{fig:gen_sol_552_0.50}}
\caption{Example minimum-energy kink solutions for the vacuum configuration of (\ref{eq:gen_rep}, \ref{eq:gen_sol_matrix}), for a selection of scalar mass (in GeV) and $\tan\beta$ values. Inset plots show the quantities $\partial_x\eta_2$ and $\partial_x\eta_3$, which converge to zero or otherwise equivalent profile for each solution.}
\label{fig:gen_sol_set}
\end{figure*}

Here we present a selection of additional minimum-energy solutions for various parameter sets in Fig.~\ref{fig:gen_sol_set}.

\end{document}